\documentclass[12pt]{iopart}
\usepackage{graphicx}
\usepackage{amssymb}
\usepackage{amsfonts}
\usepackage{url}
\usepackage{cite}

\usepackage{epstopdf}
\usepackage{color}
\usepackage{bm}
\usepackage[colorlinks=true,linkcolor=blue,citecolor=red]{hyperref}
\usepackage{comment}

\newcommand{\binom}[2]{{#1 \choose #2}}

\def\nnu{\mu}

\def\qc{q_{\rm c}}
\def\qccrit{{q_{\rm c}^{\rm crit}}}

\def\K{\Gamma}
\def\L{\mathcal L}
\def\M{\mathcal M}
\def\N{\mathcal N}

\def\pa{\partial\Omega}
\def\E{{\mathbb E}}
\def\P{{\mathbb P}}
\def\R{{\mathbb R}}

\def\erfc{\mathrm{erfc}}
\def\erfcx{\mathrm{erfcx}}

\def\x{\bm{x}}

\begin{document}

\title{Diffusion-Driven Surface Autocatalysis in Exterior Domains}

\author{Denis~S.~Grebenkov}
 \ead{denis.grebenkov@polytechnique.edu}
\address{
Laboratoire de Physique de la Mati\`{e}re Condens\'{e}e, \\ 
CNRS -- Ecole Polytechnique, Institut Polytechnique de Paris, 91120 Palaiseau, France}

\date{\today}

\begin{abstract}
We study the collective dynamics of independent particles that diffuse
outside a spherical surface, on which they are replicated with a
prescribed catalytic rate.  In spatial dimensions three and higher,
the transient nature of diffusion creates a competition between
autocatalytic branching and escape to infinity, leading to three
distinct regimes: subcritical (finite limiting population), critical
(algebraic growth), and supercritical (exponential growth).  The
rotational symmetry of the domain and an explicit form of the
single-particle diffusion propagator allow us to fully investigate the
time evolution of the population size, including the mean population
size, variance, higher-order moments, as well as the whole
distribution.  In particular, we obtain an explicit form of the
distribution at long times and describe a slow, power-law approach to
this steady-state limit.  We further extend the steady-state analysis
to general exterior domains using the Dirichlet-to-Neumann operator,
which determines the critical catalytic rate and reveals how boundary
geometry affects the population-size distribution.
\end{abstract}





\submitto{\JPA}

\maketitle

\section{Introduction}

Diffusion-reaction processes are ubiquitous in nature and industrial
applications, with examples ranging from metabolism and gene
regulation in cells to heterogeneous catalysis
\cite{Weibel,Ben-Avraham,Murrey,Krapivsky,Bressloff13,Lindenberg,Grebenkov}.
From the theoretical point of view, two mainstream directions have
been particularly well studied: autocatalytic reactions in the bulk
and diffusion-controlled reactions at surfaces.  In the first setting,
each particle diffusing in the bulk may spontaneously branch (split)
into two or many copies of itself \cite{Athreya,Kolmogorov47}.  The
growth of the population of particles can be controlled by eventual
killing mechanisms in the bulk or at the surface
\cite{Sevastyanov58,Skorohod64}.  A probabilistic description of such
branching processes relies on the theory of measure-valued processes
(or superprocesses) \cite{LeGall,Dynkin}, whereas their generating
functions and related observables are often described by nonlinear
partial differential equations (PDEs), e.g., the Fisher-KPP equation
\cite{Fisher37,Kolmogorov37,Grindrod}.  In the second setting,
diffusing particles react or disappear at a surface, with examples
including heterogeneous catalysis, recombination events, spin
relaxation on magnetic impurities, oxygen uptake, and various
biochemical reactions inside living cells
\cite{North66,Wilemski73,Sano79,Brownstein79,Calef83,Berg85,Grebenkov07,Serov16,Galanti16,Grebenkov23g}.
The kinetics of such diffusion-controlled reactions are usually
described by {\it linear} PDEs or integral equations
\cite{Redner,Schuss,Metzler}.  In this case, one aims at understanding
how the geometric complexity of the environment and reactive
properties of the surface affect the survival probability of a
particle or, equivalently, the associated distribution of its
first-passage or first-reaction times
\cite{Holcman13,Benichou14,Lanoiselee18}.   

A natural intersection of these two fields corresponds to
autocatalytic reactions at a surface.  In the mathematical literature,
such reactions are generally introduced by considering Brownian motion
with heterogeneous branching on a lower-dimensional subset in $\R^d$
\cite{Dawson99,Klenke00,Kesten03,Englander04,Morters05,Englander07} or
lattice random walks with a finite number of catalytic sites
\cite{Redner84,Albeverio98,Bogachev98,Vatutin04,Vatutin07,Hu12,Topchi13,Bocharov14,Bulinskaya18,Bauer21}.
While both approaches offer significant flexibility and generality,
the geometric structure of the environment appears less explicitly.
In order to understand the role of the geometry in the autocatalytic
dynamics, it is therefore more convenient to rely on integral
equations or PDE-based descriptions
\cite{DelGrosso76,Delmas05,Barbu16}.  Recently, we proposed a generic
model of autocatalytic reactions at surfaces and provided its
theoretical description via a linear diffusion equation with {\it
nonlinear} Robin-type boundary conditions
\cite{Grebenkov26b,Grebenkov26c}.  More precisely, we studied the
population size $\N(t)$, i.e., the number of particles present at time
$t$, resulting from branching events on catalytic regions of the
boundary and eventual killing mechanisms on other boundary regions or
in the bulk.  The distribution of this random variable is fully
characterized by its generating function $G_s(t|\x_0) = \E_{\x_0}\{
s^{\N(t)}\}$, where $\x_0$ is the starting point of the first particle
at time $0$, and $s \in [0,1]$ is a parameter.  We established several
equivalent representations of the generating function in terms of the
single-particle propagator.  For bounded domains, the long-time
asymptotic behavior of $G_s(t|\x_0)$ was analyzed in the subcritical,
critical, and supercritical regimes when the mean population size
vanishes, reaches a steady-state limit, or grows exponentially,
respectively.

In this paper, we extend the above analysis to unbounded domains by
applying the developed framework to an archetypal model of diffusion
outside a ball with autocatalytic reactions at its spherical boundary.
Although this model does not include killing mechanisms, the transient
character of diffusion in $\R^d$ with $d\geq 3$ introduces the
possibility of escape to infinity.  The competition between branching
and escape leads to distinct subcritical, critical, and supercritical
regimes, with population-size statistics that differ significantly
from those in bounded domains.  Exploiting the explicit
single-particle propagator for diffusion outside a ball, we derive the
long-time behavior of the moments and obtain an explicit steady-state
distribution of the population size, despite the nonlinear nature of
the underlying phenomenon.  In the supercritical regime, we also
investigate the limiting distribution of the exponentially rescaled
population.  Finally, we extend the steady-state analysis to general
exterior domains using the Dirichlet-to-Neumann operator, which
provides a geometric characterization of the critical catalytic rate
and a framework for studying the large-population asymptotics.

The paper is organized as follows.  In Sec. \ref{sec:model}, we
describe the model and recall the theoretical framework established in
\cite{Grebenkov26b}.  Section \ref{sec:3D} presents our main results
for a spherical surface in three dimensions.  In
Sec. \ref{sec:higher}, we discuss the asymptotic behavior of the mean
population size in higher dimensions and compare it to lattice-based
models.  Section \ref{sec:generic_domains} provides an extension to
generic exterior domains, while Sec. \ref{sec:conclusion} concludes
the paper.

\section{Theoretical description of autocatalytic dynamics}
\label{sec:model}

We consider diffusion in a domain $\Omega \subset \R^d$, which is the
exterior of a compact set, with sufficiently smooth boundary $\pa$.
At time $t = 0$, a single point-like particle is released from a point
$\x_0 \in \Omega$ and starts diffusing in $\Omega$ with a constant
diffusivity $D > 0$.  Upon encountering the catalytic boundary $\pa$,
the particle can be either reflected back to continue bulk diffusion,
or be transformed into two identical copies of itself.  In the latter
case, two newborn particles diffuse independently of each other.  The
(infinitely small) probability of the branching event is proportional
to a catalytic rate $\qc > 0$ and is controlled by the local time
spent on the boundary.  More rigorously, the catalytic boundary $\pa$
keeps reflecting the particle, until its boundary local time exceeds
an independent random variable $\hat{\ell}$ with the exponential
distribution $\P\{\hat{\ell} > \ell\} = e^{-\qc \ell}$ (see
\cite{Grebenkov26c,Grebenkov20} for more details on the probabilistic
description of catalytic events driven by the boundary local time).
Even though $\qc$ has units of inverse length, we keep calling it
catalytic ``rate''.

In two dimensions, the recurrent nature of planar Brownian motion
ensures that each particle keeps returning to the catalytic boundary
and thus will necessarily branch.  As there is no absorption event,
the particles cannot disappear, and the population of particles grows
over time and diverges as $t\to \infty$ for any $\qc > 0$.  In
contrast, for $d \geq 3$, each newborn particle has a nonzero
probability to escape to infinity and thus to never branch.  One can
therefore intuitively expect the existence of a critical catalytic
rate $\qccrit$ that distinguishes three regimes \cite{Grebenkov26a}.
(i) If $\qc < \qccrit$ (subcritical regime), branching events are not
frequent enough so that the mean population size reaches a finite
limit at long times.  (ii) If $\qc > \qccrit$ (supercritical regime),
branching events are sufficiently frequent for the expected number of
particles to grow without bound, despite the possibility of escape to
infinity.  (iii) The critical regime $\qc = \qccrit$ separates these
two settings.  This intuitive picture is mathematically justified by
the spectral analysis of the Robin Laplacian in exterior domains
\cite{Bundrock25} (see also \cite{Freitas15,Krejcirik18,Krejcirik20}).
We aim to obtain a {\it quantitative} description of these regimes for
diffusion in the exterior of a ball (see
Sec. \ref{sec:generic_domains} for generic domains).

A mathematical characterization of the population size $\N(t)$ that
was established in \cite{Grebenkov26b,Grebenkov26c} for bounded
domains, remains valid for exterior domains.  
For completeness, we briefly recall the derivation of the renewal
equation for the generating function $G_s(t|\x_0) = \E_{\x_0}\{
s^{\N(t)}\}$ of the population size $\N(t)$, with $0 \leq s \leq 1$.
First, we introduce the random time $\tau$ of the first branching
event, whose distribution is characterized by the single-particle
propagator $P^+(\x,t|\x_0)$ that obeys for any $\x_0
\in \Omega$:
\numparts
\begin{eqnarray}  \label{eq:Pplus_diff}
&& \partial_t P^+(\x,t|\x_0) - D \Delta P^+(\x,t|\x_0) =  0 \qquad (\x \in \Omega), \\  \label{eq:Pplus_BC}
&& \partial_n P^+(\x,t|\x_0) + \qc P^+(\x,t|\x_0) =  0 \qquad (\x \in \pa), \\
&& P^+(\x,t|\x_0) \to 0  \qquad (|\x| \to \infty), \\  \label{eq:Pplus_initial}
&& P^+(\x,0|\x_0) =  \delta(\x-\x_0),  
\end{eqnarray}
\endnumparts
where $\Delta$ is the Laplace operator, $\partial_n$ is the normal
derivative oriented outward from the domain $\Omega$, and
$\delta(\x-\x_0)$ is the Dirac distribution that fixes the initial
position $\x_0$ of the particle at $t = 0$.
The superscript plus in $P^+(\x,t|\x_0)$ highlights the positive sign
in front of $\qc$ in the Robin boundary condition (\ref{eq:Pplus_BC}),
i.e., the boundary $\pa$ is treated here as partially absorbing.  In
other words, $P^+(\x,t|\x_0)$ is the conventional heat kernel with
Robin boundary condition that admits the standard probabilistic
interpretation: $P^+(\x,t|\x_0)d\x$ is the probability of finding the
particle in $d\x$ vicinity of $\x$ at time $t$, given that it has
started at $\x_0$ at time $0$ and has not reacted on $\pa$ up to time
$t$.  For a sufficiently smooth boundary, the propagator also admits
starting from the boundary, $\x_0\in\pa$, and satisfies the symmetry
relation $P^+(\x,t|\x_0)=P^+(\x_0,t|\x)$.

Then, the generating function can be decomposed into the sum of
contributions from two complementary events:
\begin{equation}  \label{eq:Gs_integral}
\fl
G_s(t|\x_0) = s \, \int\limits_{\Omega} d\x \, P^+(\x,t|\x_0) 
+ \int\limits_0^t dt'\, \int\limits_{\pa} d\x' \, \qc D P^+(\x',t'|\x_0) \, G_s^2(t-t'|\x').
\end{equation}
The first term represents the probability of no branching up to time
$t$, $\P_{\x_0}\{ \tau > t\}$, in which case $\N(t) = 1$ that produces
the factor $s^1$.  In turn, if the branching occurred before $t$, the
probability flux density, $-D\partial_n P^+(\x',t'|\x_0) = \qc D
P^+(\x',t'|\x_0)$, determines the joint probability density of the
branching position $\x'\in \pa$ and the branching time $t'$.  Upon the
branching event, two particles are released from $\x'$ and generate
{\it independently} two subpopulations during the remaining time
$t-t'$.  Averaging over all possible times and locations of the first
branching event yields the second term.  The extension to branching
into $m$ descendants is immediate: one replaces $G_s^2(t-t'|\x')$ by
$G_s^m(t-t'|\x')$ in Eq. (\ref{eq:Gs_integral}).  Compared to
Refs. \cite{Grebenkov26b,Grebenkov26c}, the present setting has no
absorbing region and allows a nonzero probability that the initial
particle never branches, corresponding to $\tau=\infty$.

By definition, the generating function $G_s(t|\x_0)$ determines the
distribution of $\N(t)$ and its moments of positive-integer orders:
\begin{eqnarray} \label{eq:Qk_def}
Q_k(t|\x_0) & =& \P_{\x_0}\{ \N(t) = k \} = \frac{1}{k!} \lim\limits_{s\to 0} \partial_s^k G_s(t|\x_0)  \qquad (k = 1,2,\ldots),\\ \label{eq:Nk_def}
N_k(t|\x_0) & =& \E_{\x_0}\{ [\N(t)]^k \} = \lim\limits_{s\to 1} (s\partial_s)^k G_s(t|\x_0)  \qquad (k = 1,2,\ldots).
\end{eqnarray}
Evaluating the $k$-th order derivative of Eq. (\ref{eq:Gs_integral})
at $s=0$ gives the integral equation for the probability
$Q_k(t|\x_0)$:
\begin{equation} \label{eq:Qn_integral}
\fl
Q_k(t|\x_0)  = \delta_{k,1} \int\limits_{\Omega} d\x \, P^{+}(\x,t|\x_0) 
+ \qc D \int\limits_0^t dt'  \int\limits_{\pa} d\x \, P^{+}(\x,t'|\x_0) \, H_k(t-t'|\x) ,
\end{equation}
where
\begin{equation}  \label{eq:Hn_def}
H_k(t|\x) = \sum\limits_{j=1}^{k-1} Q_j(t|\x) Q_{k-j}(t|\x)
\end{equation}
is obtained by using the generalized Leibniz rule upon evaluating the
derivatives of $G_s^2$.  Since there is no absorbing region, the
particle cannot disappear so that $Q_0(t|\x_0) \equiv 0$ is excluded
from the sum in Eq. (\ref{eq:Hn_def}).  Similarly, according to
Eq. (\ref{eq:Nk_def}), one can apply the operator $(s\partial_s)^k$ to
Eq. (\ref{eq:Gs_integral}) to deduce the integral equation for the
$k$-th moment $N_k(t|\x_0)$:
\begin{equation}  \label{eq:Nk_Fk}
\fl
N_k(t|\x_0) = \int\limits_{\Omega} d\x \, P^-(\x,t'|\x_0) 
+\qc D \int\limits_0^t dt' \int\limits_{\pa} d\x  \, P^-(\x,t'|\x_0) \, F_k(t-t'|\x),
\end{equation}
where
\begin{equation}  \label{eq:Fk_def}
F_k(t|\x) = \sum\limits_{j=1}^{k-1} \binom{k}{j} N_j(t|\x) \, N_{k-j}(t|\x)
\end{equation}
(see details in Refs. \cite{Grebenkov26b,Grebenkov26c}).  For
convenience, this integral equation employs another single-particle
propagator $P^-(\x,t|\x_0)$ satisfying
\numparts
\begin{eqnarray}
&& \partial_t P^-(\x,t|\x_0) - D \Delta P^-(\x,t|\x_0) = 0 \qquad (\x \in \Omega), \\
&& \partial_n P^-(\x,t|\x_0) - \qc P^-(\x,t|\x_0) = 0 \qquad (\x \in \pa), \\
&& P^-(\x,t|\x_0) \to 0 \qquad (|\x| \to \infty), \\
&& P^-(\x,0|\x_0) = \delta(\x-\x_0) . 
\end{eqnarray}
\endnumparts
Here, the sign in front of $\qc$ is negative, i.e., the boundary $\pa$
is treated as catalytic.  As discussed in \cite{Grebenkov26c},
$P^-(\x,t|\x_0)d\x$ represents the expected number of particles
located in an infinitesimal volume $d\x$ around $\x$ at time $t$,
given that the initial particle started at $\x_0$.
Even though the integral equations (\ref{eq:Qn_integral},
\ref{eq:Nk_Fk}) can be recast in a PDE form
(see \cite{Grebenkov26b,Grebenkov26c}), we keep using their integral
form in this paper.

In the next Section, we will exploit the explicit form of the
propagators $P^\pm(\x,t|\x_0)$ for the exterior of a ball to analyze
in depth the properties of $Q_k(t|\x_0)$ and $N_k(t|\x_0)$.

\section{Autocatalysis at a spherical surface in three dimensions}
\label{sec:3D}

Throughout this section, we focus on the exterior of a ball of radius
$R$ in three dimensions: $\Omega = \{\x\in\R^3 ~:~ |\x|> R\}$.

\subsection{Propagators}

The key advantage of the considered geometric setting is the explicit
form of the propagators $P^{\pm}(\x,t|\x_0)$ for the exterior of a
ball.  Moreover, the rotational symmetry of this domain implies that
both $Q_k(t|\x_0)$ and $N_k(t|\x_0)$ are rotationally invariant, i.e.,
their spatial dependence is reduced to the radial coordinate $r_0 =
|\x_0|$.  As a consequence, one can employ the spherical coordinates,
$\x_0 = (r_0,\theta_0,\varphi_0)$, and average the propagator over
angular coordinates $\theta_0$ and $\varphi_0$ to get (see, e.g.,
Eqs. (RS20, RS23) from \cite{Cole}):
\begin{eqnarray} \nonumber
\fl
P^+(r,t|r_0) &=& \frac{1}{4\pi} \int\limits_0^\pi d\theta \, \sin\theta_0 \int\limits_0^{2\pi} d\varphi_0 \, P^+(\x,t|\x_0) 
= \frac{\exp\bigl(-\frac{(r-r_0)^2}{4Dt}\bigr) + \exp\bigl(-\frac{(r+r_0-2R)^2}{4Dt}\bigr)}{8\pi r r_0 \sqrt{\pi Dt}} \\  \label{eq:P+}
\fl &-& \frac{h_+}{4\pi rr_0 R} \exp\left(-\frac{(r+r_0-2R)^2}{4Dt}\right)
\erfcx\left(\frac{r+r_0-2R}{\sqrt{4Dt}} + h_+ \frac{\sqrt{Dt}}{R}\right), 
\end{eqnarray} 
where $h_+ = 1 + \qc R$, $r = |\x|$, and $\erfcx(z) = e^{z^2}
\erfc(z)$ is the scaled complementary error function.
We used the shortcut notation $P^+(r,t|r_0)$ to highlight the
dependence on radial coordinates.  In the special case $r = r_0 = R$,
the above relation is simplified to
\begin{equation}   \label{eq:P+R}
P^+(R,t|R) = \frac{1}{4\pi R^3} \left[\frac{R}{\sqrt{\pi Dt}} - h_+ \, \erfcx \left(h_+ \sqrt{Dt}/R\right)\right].
\end{equation}  
The same expressions hold for $P^-(r,t|r_0)$ and $P^-(R,t|R)$ if $h_+$
is replaced by $h_- = 1 - \qc R$.  In particular, if $\qc = 1/R$, one
gets a very simple form $P^-(R,t|R) = 1/(4\pi R^2 \sqrt{\pi Dt})$ that
also determines the short-time behavior of the propagators
$P^{\pm}(R,t|R)$ for any $\qc$:
\begin{equation}
P^{\pm}(R,t|R) \simeq \frac{1}{4\pi R^2 \sqrt{\pi Dt}} \qquad (t\to 0).
\end{equation} 
In contrast, the long-time behavior depends on the sign of $h_{\pm}$:
if $h_{\pm} > 0$, one uses $\erfcx(z) \simeq (1 - 1/(2z^2) +
\ldots)/(z\sqrt{\pi})$ as $z\to \infty$ to show that
\begin{equation}
P^{\pm}(R,t|R) \simeq \frac{1}{h_\pm^2 (4\pi Dt)^{3/2}} \qquad (t\to \infty).
\end{equation} 
In contrast, when $h_- < 0$, one has $\erfcx(z) \simeq 2e^{z^2}$ so
that the propagator $P^-(R,t|R)$ diverges exponentially fast as
$t\to\infty$ (this is not possible for $P^+(R,t|R)$, for which $h_+
\geq 1$).

\subsection{Mean population size}

The mean population size $N_1(t|\x_0)$ follows directly by setting $k
= 1$ to Eq. (\ref{eq:Nk_Fk}), realizing that $F_1 \equiv 0$, and
integrating the propagator $P^-(\x,t|\x_0)$ over the exterior of the
ball.  The explicit formula for $\qc R \ne 1$ was given in
\cite{Grebenkov26a}:
%
\begin{eqnarray} \nonumber
\fl
N_1(t|\x_0) &=& 1 - \frac{R/r_0}{1-1/(\qc R)} e^{-(r_0-R)^2/(4Dt)} \\  \label{eq:N1}
\fl
&\times& \left\{ \erfcx\left(\frac{r_0-R}{\sqrt{4Dt}}\right) 
- \erfcx\left(\frac{r_0-R}{\sqrt{4Dt}} + \sqrt{Dt}(-\qc + 1/R)\right) \right\}.
\end{eqnarray}
At $r_0 = R$, we get 
\begin{equation} 
N_1(t|R) = 1 - \frac{1}{1-1/(\qc R)}  \biggl\{ 1 - \erfcx\biggl(\sqrt{Dt}(-\qc + 1/R)\biggr) \biggr\}.
\end{equation}
Using the asymptotic behavior of $\erfcx(z)$ as $z\to \infty$, one
sees that the critical value $\qccrit = 1/R$ distinguishes three
asymptotic regimes:

(i) In the subcritical regime ($\qc < \qccrit$), the mean population
size approaches a constant limit:
\begin{equation}  \label{eq:N1_subcrit}
N_1(t|\x_0) \to N_1(\infty|\x_0) = 1 + \frac{R/r_0}{1/(\qc R) - 1} \,.
\end{equation}

(ii) In the critical regime ($\qc = \qccrit$), one has
\begin{equation}  \label{eq:N1_crit}
N_1(t|\x_0) = 1 + \frac{2\sqrt{Dt}}{\sqrt{\pi} r_0} e^{-(r_0-R)^2/(4Dt)} 
 - (1 - R/r_0)\, \erfc\left(\frac{r_0-R}{\sqrt{4Dt}}\right) ,
\end{equation}
which exhibits a power-law growth at long times:
\begin{equation}
N_1(t|\x_0) \simeq 1 + \frac{2\sqrt{Dt}}{\sqrt{\pi} r_0} \qquad (t\to \infty).
\end{equation}

(iii) In the supercritical regime ($\qc > \qccrit$), the mean
population size grows exponentially with time:
\begin{equation}  \label{eq:N1_supercrit}
N_1(t|\x_0) \simeq \frac{2R/r_0}{1 - 1/(\qc R)}\, e^{\mu t}  \qquad (t\to \infty),
\end{equation}
with the rate 
\begin{equation} \label{eq:mu_def} 
\nnu = D(\qc - 1/R)^2.
\end{equation}

\subsection{Higher-order moments}
\label{sec:Nk}

For a catalytic sphere, the integral equation (\ref{eq:Nk_Fk}) is
reduced to
\begin{equation}  \label{eq:Nk_int}
N_k(t|r_0) = N_1(t|r_0) + 4\pi R^2 \qc D \int\limits_0^t dt' \, P^-(R,t'|r_0)\, F_k(t-t'|R).
\end{equation}
Setting $r_0 = R$ and using Eq. (\ref{eq:Fk_def}), we have
\begin{equation}  \label{eq:Nk_intR}
\fl
N_k(t|R) = N_1(t|R) + 4\pi R^2 \qc D \int\limits_0^t dt' \, P^-(R,t-t'|R)\, 
\sum\limits_{j=1}^{k-1} \binom{k}{j} N_j(t'|R) N_{k-j}(t'|R).
\end{equation}
As the right-hand side depends on $N_1(t|R),\cdots,N_{k-1}(t|R)$, this
equation can be solved recursively to determine $N_k(t|R)$ (see
\ref{sec:numerics} for its numerical solution).

We inspect separately three asymptotic regimes.

\subsubsection*{Critical regime.}

In the critical regime, Eq. (\ref{eq:N1_crit}) is reduced to
\begin{equation}
N_1(t|R) = 1 + \frac{2\sqrt{Dt}}{\sqrt{\pi} R} \,,
\end{equation}
whereas Eq. (\ref{eq:P+R}), written for $P^-(R,t|R)$, yields 
\begin{equation}
P^-(R,t|R) = \frac{1}{4\pi R^2 \sqrt{\pi Dt}} \,.
\end{equation}
As a consequence, the integral equation (\ref{eq:Nk_intR}) can be
solved explicitly for several $k$; e.g.,
\begin{equation}  \label{eq:N2}
N_2(t|R) = 1 + \frac{4\sqrt{Dt}}{\sqrt{\pi} R} \left(\frac32 + \frac{\sqrt{\pi Dt}}{R} + \frac{8Dt}{3\pi R^2}\right).
\end{equation}
One sees that $N_2(t|R) \propto t^{3/2}$ at long times.  In
particular, the coefficient of variation, $\gamma(t) =
\bigl[N_2(t|R)/N_1^2(t|R) - 1\bigr]^{1/2}$, that characterizes the
broadness of the distribution with respect to its mean, grows as
$t^{1/4}$.

\begin{figure}
\begin{center}
\includegraphics[width=0.32\textwidth]{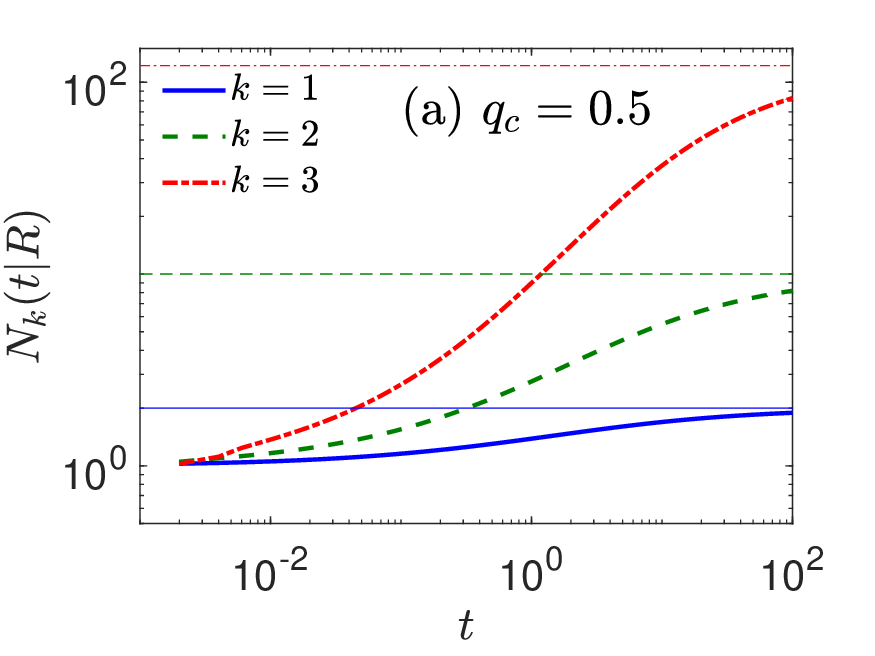} 
\includegraphics[width=0.32\textwidth]{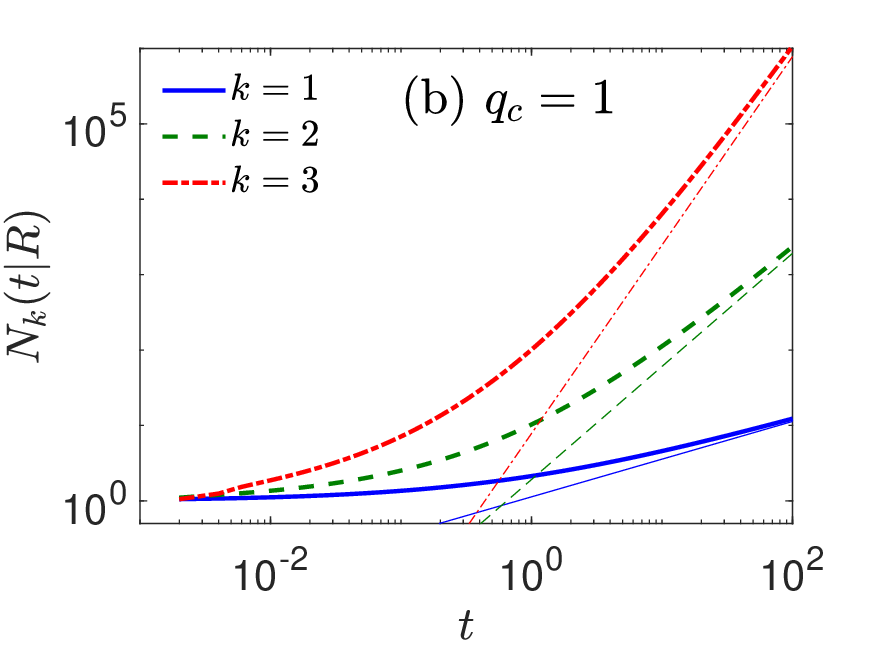} 
\includegraphics[width=0.32\textwidth]{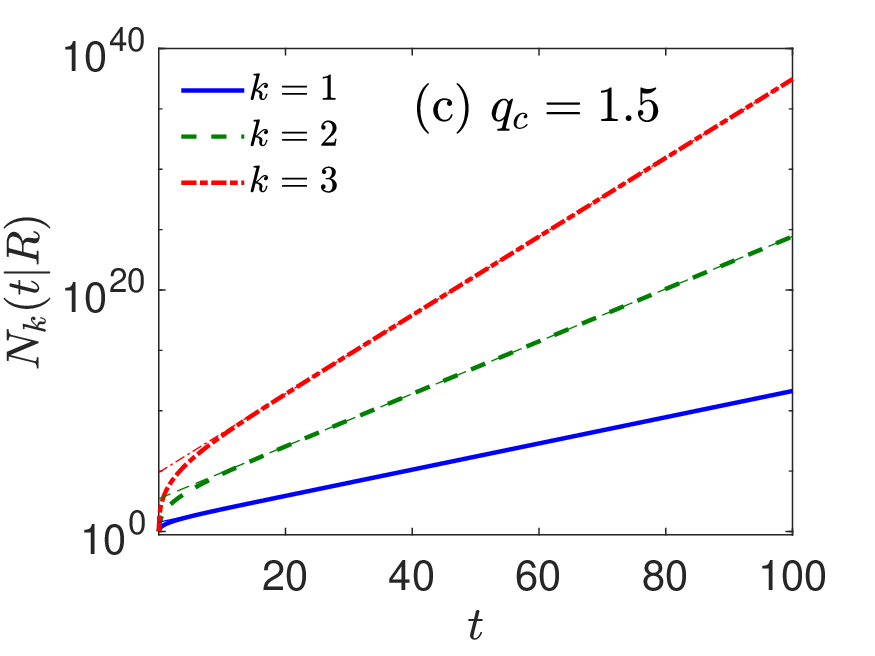} 
\end{center}
\caption{
Moments $N_k(t|R)$ of the population size for diffusion with $D = 1$
outside the unit ball ($R = 1$) in the subcritical ($\qc = 0.5$, panel
a), critical ($\qc = 1$, panel b), and supercritical ($\qc = 1.5$,
panel c) regimes.  Thick lines present the numerical integration of
Eqs. (\ref{eq:Nk_intR}), whereas thin lines show the asymptotic
behavior: Eq. (\ref{eq:Nk_asympt_subcrit}) for panel a,
Eq. (\ref{eq:Nk_asympt_crit}) for panel b, and
Eq. (\ref{eq:Nk_supercrit}) for panel c. }
\label{fig:Nk}
\end{figure}

In general, one can use the induction argument and evaluate the
convolution in Eq. (\ref{eq:Nk_intR}) to deduce the leading-order
long-time behavior:
\begin{equation}  \label{eq:Nk_asympt_crit}
N_k(t|R) \propto n_k \, (Dt/R^2)^{k-1/2}  \qquad (t\to \infty) ,
\end{equation}
where the prefactors $n_k$ are determined recursively:
\begin{equation}
n_1 = \frac{2}{\sqrt{\pi}} \,, \qquad 
n_k = \frac{\Gamma(k)}{\Gamma(k+1/2)} \sum\limits_{j=1}^{k-1} \binom{k}{j} n_j \, n_{k-j} \,.
\end{equation}
For instance, one gets $n_2 = 32/(3\pi^{3/2})$, in agreement with the
leading-order term in the explicit solution (\ref{eq:N2}).
The panel b of Fig. \ref{fig:Nk} illustrates the behavior of the
first three moments $N_k(t|R)$ for the unit ball that were obtained
via a numerical integration of Eqs. (\ref{eq:Nk_intR}), see
\ref{sec:numerics} for details.  The leading-order relation
(\ref{eq:Nk_asympt_crit}) accurately describes the long-time
asymptotic behavior.

\subsubsection*{Subcritical regime.}

For the analysis of the subcritical regime, it is instructive to
evaluate the integral
\begin{equation}  \label{eq:Pint}
\int\limits_0^t dt' \, P^\pm(R,t'|R) = \frac{1 - \erfcx(h_\pm\sqrt{Dt}/R)}{4\pi RD h_\pm} \,.
\end{equation}
For $h_{\pm} = 1 \pm \qc R > 0$, this integral approaches the constant
$1/(4\pi RD h_\pm)$ as $t\to \infty$.

Setting
\begin{equation}  \label{eq:gammam}
\gamma_- = \frac{\qc R}{|1 - \qc R|} 
\end{equation}
into Eq. (\ref{eq:N1_subcrit}), we have $N_1(\infty|R) = 1 +
\gamma_-$.  The other moments also reach steady-state limits.  Using
the induction argument, we determine these limits by substituting the
limit of Eq. (\ref{eq:Pint}) into Eq. (\ref{eq:Nk_intR}):
\begin{equation}  \label{eq:Nk_asympt_subcrit}
N_k(\infty|R) = 1 + \gamma_- \left(1 + \sum\limits_{j=1}^{k-1} \binom{k}{j} N_j(\infty|R) N_{k-j}(\infty|R)\right).
\end{equation}
For instance, we get
\begin{eqnarray*}
N_1(\infty|R) & =& 1 + \gamma_- , \\
N_2(\infty|R) & =& 1 + \gamma_-  + 2\gamma_- [N_1(\infty|R)]^2, \\
N_3(\infty|R) & =& 1 + \gamma_-  + 6\gamma_- N_1(\infty|R) N_2(\infty|R) .
\end{eqnarray*}
Evaluating explicitly the generating function of $N_k(\infty|R)$
and using standard combinatorial analysis, one can deduce the
following nonrecursive representation:
\begin{equation}
N_k(\infty|R) = (1+\gamma_-) \sum\limits_{j=0}^{k-1} \left\{ k \atop {j+1} \right\} \, 2^j (2j-1)!!\, [\gamma_-(1+\gamma_-)]^j ,
\end{equation}
where $(-1)!! = 1$, and $\bigl\{\left. k \atop j\right. \bigr\}$ are
the Stirling numbers of the second kind, defined as $\bigl\{\left. k
\atop j\right. \bigr\} = \frac{1}{j!} \sum\nolimits_{l=1}^j (-1)^{j-l}
\binom{j}{l} l^k$.
The panel a of Fig. \ref{fig:Nk} illustrates how $N_k(t|R)$ approach
these limiting values.

\subsubsection*{Supercritical regime.}

In the supercritical regime ($\qc > 1/R$), Eq. (\ref{eq:N1_supercrit})
states that the mean population size grows exponentially at long
times, with the rate $\nnu$ given by Eq. (\ref{eq:mu_def}).
Similarly, the propagator $P^-(R,t|R)$ behaves as
\begin{equation}
P^-(R,t|R) \simeq \frac{\qc R-1}{2\pi R^3} e^{\nnu t}  \quad (t\to\infty).
\end{equation}

We aim to show that 
\begin{equation}    \label{eq:Nk_supercrit}
N_j(t|R) \simeq n_j\, e^{j\nnu t} \qquad (t\to\infty),
\end{equation}
with some coefficients $n_j$.  For this purpose, we assume that this
relation holds for $j=1,\ldots,k-1$ and then use the induction
argument by checking its validity for $k$.  To eliminate the
convolution form in Eq. (\ref{eq:Nk_int}), it is convenient to apply
the Laplace transform:
\begin{equation}  \label{eq:Nk_int_Laplace}
\tilde{N}_k(p|r_0) = \tilde{N}_1(p|r_0) + 4\pi R^2 \qc D \tilde{P}^-(R,p|r_0)\, \tilde{F}_k(p|R),
\end{equation}
where tilde denotes the Laplace transform, e.g., $\tilde{N}_k(p|r_0) =
\int\nolimits_0^\infty dt \, e^{-pt} \, N_k(t|r_0)$.  The long-time
behavior is determined by the largest pole in Laplace domain.  The
largest pole of $\tilde{N}_1(p|r_0)$ and $\tilde{P}^-(R,p|r_0)$ is
$\nnu$.  In turn, substituting Eq. (\ref{eq:Nk_supercrit}) with $j=
1,2,\ldots, k-1$ into Eq. (\ref{eq:Fk_def}), one has $F_k(t|R) \propto
e^{k\nnu t}$ as $t\to \infty$, so that the largest pole of
$\tilde{F}_k(p|R)$ is $k\nnu$.  As a consequence, we obtain for $k \geq
2$ as $t\to \infty$:
\begin{eqnarray*} \nonumber
N_k(t|R) & \simeq& 4\pi R^2 \qc D F_k(t|R) \underbrace{\int\limits_0^\infty dt' P^{-}(R,t'|R) e^{-k \nnu t'}}_{=\tilde{P}^-(R,k\nnu|R)} \\  \nonumber
& =& F_k(t|R) \frac{\qc R}{R\sqrt{k\nnu /D} + 1 - \qc R} \\
& \simeq & \frac{e^{k\nnu t}}{(\sqrt{k}-1)(1 - 1/(\qc R))}  \sum\limits_{j=1}^{k-1} \binom{k}{j} n_j n_{k-j} \,,
\end{eqnarray*}
where we used the Laplace transform of $P^{-}(R,t|R)$, evaluated at $p
= k\nnu$.  We conclude that Eq. (\ref{eq:Nk_supercrit}) holds for $j=
k$, with
\begin{equation}  \label{eq:nk_recursive}
n_1 = 2\gamma_- , \qquad   n_k = \frac{\gamma_-}{\sqrt{k}-1}  \sum\limits_{j=1}^{k-1} \binom{k}{j} n_j n_{k-j} \,,
\end{equation}
where $\gamma_-$ was defined in Eq. (\ref{eq:gammam}).
The panel c of Fig. \ref{fig:Nk} illustrates the accuracy of the
asymptotic relation (\ref{eq:Nk_supercrit}).

\subsection{Distribution}
\label{sec:distrib}

In the same vein, we can analyze the probabilities $Q_k(t|\x_0)$.
First, we note that the probability of no branching, $Q_1(t|\x_0)$,
corresponds to the conventional survival probability of a single
particle in the presence of a partially reactive sphere with
reactivity $\qc$; in other words, one retrieves the standard Collins
and Kimball's result \cite{Collins49}:
\begin{eqnarray} \nonumber
Q_1(t|\x_0) &=& 1 - \frac{R/r_0}{1+1/(\qc R)} e^{-(r_0-R)^2/(4Dt)} \\  \label{eq:Q1}
& \times& \left\{ \erfcx\left(\frac{r_0-R}{\sqrt{4Dt}}\right) 
- \erfcx\left(\frac{r_0-R}{\sqrt{4Dt}} + \sqrt{Dt}(\qc + 1/R)\right) \right\},
\end{eqnarray}
which also follows from Eq. (\ref{eq:N1}) by changing the sign of
$\qc$.  At long times, this probability reaches the steady-state limit
\begin{equation} 
Q_1(\infty|\x_0) = 1 - \frac{R/r_0}{1+1/(\qc R)}  \,.
\end{equation}

The other probabilities can be determined by solving the integral
equation (\ref{eq:Qn_integral}) for $k \geq 2$.  For a catalytic
sphere, it reads as
\begin{equation}  \label{eq:Qn}
Q_k(t|r_0) = 4\pi R^2 \qc D \int\limits_0^t dt' \, P^+(R,t-t'|r_0) \sum\limits_{j=1}^{k-1} Q_j(t'|R) Q_{k-j}(t'|R).
\end{equation}
Setting $r_0 = R$, we get for $k = 2,3,\ldots$
\begin{equation}  \label{eq:QnR}
Q_k(t|R) = 4\pi R^2 \qc D \int\limits_0^t dt' \, P^+(R,t-t'|R) \sum\limits_{j=1}^{k-1} Q_j(t'|R) Q_{k-j}(t'|R).
\end{equation}
Since the right-hand side depends on $Q_1, \ldots, Q_{k-1}$, one can
compute $Q_k$ recursively for $k= 2,3,\ldots$.  Moreover, as
$P^+(R,t|R)$ and $Q_1(t|R)$ are known explicitly, the numerical
computation is rather straightforward (see \ref{sec:numerics} for
details).

In the long-time limit, one can apply the argument that we used for
$N_k(t|R)$, namely, to substitute the limit of Eq. (\ref{eq:Pint}) to
Eq. (\ref{eq:Qn}) in order to get the steady-state distribution:
\begin{equation}  \label{eq:Qk_relation}
Q_k(\infty|R) = \gamma_+ \sum\limits_{j=1}^{k-1} Q_j(\infty|R) Q_{k-j}(\infty|R) \qquad (k = 2,3,\ldots),
\end{equation}
where 
\begin{equation}  \label{eq:gamma_plus}
\gamma_+ = \frac{\qc R}{1 + \qc R} \,.  
\end{equation}
Multiplying Eq. (\ref{eq:Qk_relation}) by $z^k$, summing over $k
\geq 2$, adding the term $zQ_1(\infty|R) = z(1-\gamma_+)$ to both
sides, and solving the quadratic equation, we obtain the generating
function 
\begin{equation}  \label{eq:qz_def}
q(z) = \sum\limits_{k=1}^\infty z^k Q_k(\infty|R) = \frac{1 - \sqrt{1 - 4z \gamma_+(1-\gamma_+)}}{2\gamma_+} \,,
\end{equation}
from which the steady-state probabilities $Q_k(\infty|R)$ follow.
Moreover, since Eq. (\ref{eq:Qk_relation}) resembles the definition of
Catalan numbers:
\begin{equation}
C_0 = 1, \qquad C_k = \sum\limits_{j=1}^k C_{j-1} C_{k-j} ,
\end{equation}
one gets
\begin{equation}  \label{eq:Qk_asympt}
Q_k(\infty|R) = C_{k-1} (1-\gamma_+) [\gamma_+(1-\gamma_+)]^{k-1} \qquad (k = 1,2,\ldots).
\end{equation}
Remarkably, this limit is valid for any regime.  This is one of the
main results of the paper.  Note that the Catalan numbers admit an
explicit form
\begin{equation}  \label{eq:Catalan}
C_{k-1} = \frac{1}{k} \binom{2k-2}{k-1}  \qquad (k=1,2,\ldots).
\end{equation}

The generating function in Eq. (\ref{eq:qz_def}) yields the
normalization:
\begin{equation}
q(1) = \sum\limits_{k=1}^\infty Q_k(\infty|R) = \frac{1 - |1 - 2\gamma_+|}{2\gamma_+} \,.
\end{equation}  
If $\qc \leq \qccrit = 1/R$ and thus $\gamma_+ \leq 1/2$, this sum is
equal to $1$, as expected.  In turn, in the supercritical regime ($\qc
> 1/R$), the right-hand side is $(1-\gamma_+)/\gamma_+ = 1/(\qc R) <
1$ because the limiting population has the nonzero probability $1-
1/(\qc R)$ of being infinite.  In this case, all moments are infinite
at $t = \infty$, in agreement with our earlier results.

Using the asymptotic behavior of Catalan numbers, one gets
\begin{equation}  \label{eq:Qkinf_largek}
Q_k(\infty|R) \simeq \frac{\bigl[4\gamma_+(1-\gamma_+)\bigr]^k}{4 \gamma_+ \sqrt{\pi}\, k^{3/2}} \qquad (k\to \infty) \,.
\end{equation}
Since $0 < \gamma_+ < 1$, one has $4\gamma_+(1-\gamma_+) \leq 1$.  In
the subcritical regime ($0 < \gamma_+ < 1/2$), the steady-state
probabilities exhibit an exponential decay with $k$ (with a power-law
prefactor $k^{-3/2}$).  In fact, the branching events are not frequent
so that particles have enough time to diffuse away from the catalytic
surface that diminishes their chances to encounter it again.  To
achieve a large population size, many generated particles have to
remain close to the catalytic boundary, and the probability of such
configurations decreases exponentially with $k$.
Interestingly, the exponential decay also occurs in the supercritical
regime $(1/2 < \gamma_+ < 1$).  Although the population grows without
bound with a nonzero probability, a finite population remains possible
because each particle may escape to infinity without further
branching.  However, trajectories involving many branching events that
nevertheless lead to a finite population become exponentially unlikely
as the population size increases.  Finally, the exponential factor
disappears in the critical regime ($\gamma_+ = 1/2$), leaving a
power-law decay $k^{-3/2}$.  This distribution is known as Catalan
distribution (or Catalan probability law) and it appears in other
branching processes (e.g., critical Galton-Watson trees).

An extension of the above results to an arbitrary starting point
$\x_0$ is straightforward.  In fact, the asymptotic limit of
Eq. (\ref{eq:Qn}) reads
\begin{equation}  \label{eq:Qn_inf}
Q_k(\infty|r_0) = 4\pi R^2 \qc D \underbrace{\int\limits_0^\infty dt' \, P^+(R,t'|r_0)}_{=1/(4\pi D r_0 h_+)} 
\underbrace{\sum\limits_{j=1}^{k-1} Q_j(\infty|R) Q_{k-j}(\infty|R)}_{=Q_k(\infty|R)/\gamma_+} ,
\end{equation}
where the integral of the propagator $P^+(R,t'|r_0)$ can be calculated
either by a direct integration of Eq. (\ref{eq:P+}) over time, or by
using the Laplace transform
\begin{equation}  \label{eq:Pp_Laplace}
\int\limits_0^\infty dt \, e^{-pt}\, P^+(R,t|r_0) = \frac{e^{-(r_0-R)\sqrt{p/D}}}{4\pi Dr_0 (1 + \qc R + R\sqrt{p/D})}
\end{equation}
and evaluating its limit as $p\to 0$.  We get thus
\begin{equation} 
Q_k(\infty|\x_0) = \left(1 - \frac{R}{r_0}\right)\delta_{k,1} + \frac{R}{r_0} \, Q_k(\infty|R) .
\end{equation}
Note that this relation could be directly obtained from a simple
probabilistic argument: the very first particle started from $\x_0$
has the probability $R/r_0$ to hit the catalytic boundary before
escaping to infinity; once the particle hits the sphere, the
probability of getting $k$ particles in the limit $t\to \infty$ is
precisely $Q_k(\infty|R)$.

In \ref{sec:Qk_long}, we further inspect the approach the steady-state
distribution and show that
\begin{equation}  \label{eq:Qk_long}
Q_k(t|R) \approx Q_k(\infty|R) + a_k \frac{R}{\sqrt{\pi Dt}} + \ldots  \qquad (t\to \infty),
\end{equation}
where
\begin{equation}  \label{eq:ak_explicit}
a_k = \frac{(2k-2)!}{[(k-1)!]^2}\, \frac{(\qc R)^{k-1}}{(1+\qc R)^{2k}} \left[\qc R - \frac{k-1}{k}\right] \qquad (k=1,2,\ldots).
\end{equation}
This implies that the leading-order correction to $Q_k(t|R)$ is
negative for $\qc R< (k-1)/k$ and positive for $\qc R > (k-1)/k$ (when
$\qc R = (k-1)/k$, one has $a_k = 0$, and a higher-order correction is
needed).  Moreover, all these corrections are positive in the critical
and supercritical regimes ($\qc R \geq 1$) so that the probabilities
$Q_k(t|R)$ are decreasing functions at large times.  Given that
$Q_k(t|R)$ for $k\geq 2$ are growing at short times, each of these
probabilities should have a maximum.

\begin{figure}
\begin{center}
\includegraphics[width=0.32\textwidth]{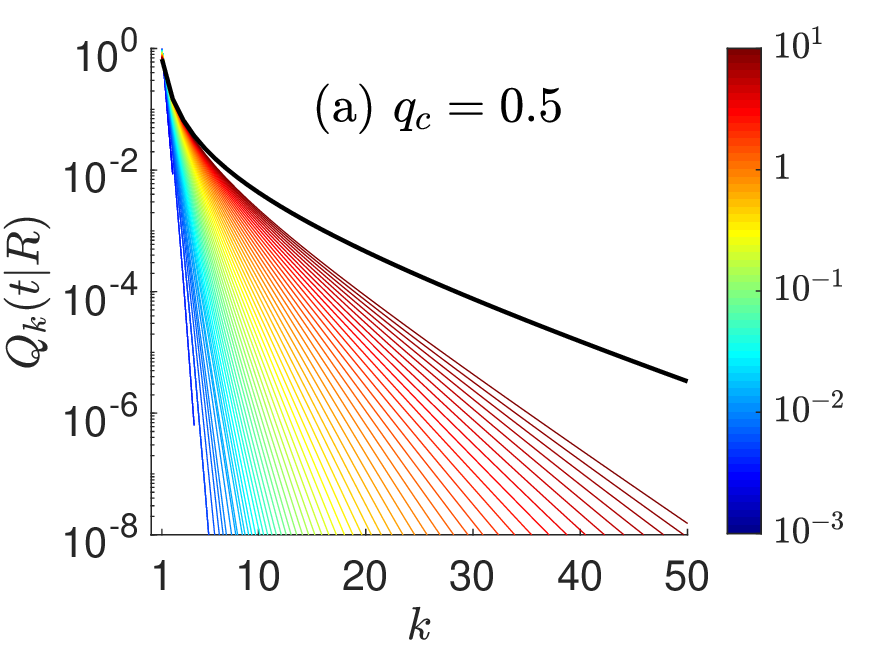} 
\includegraphics[width=0.32\textwidth]{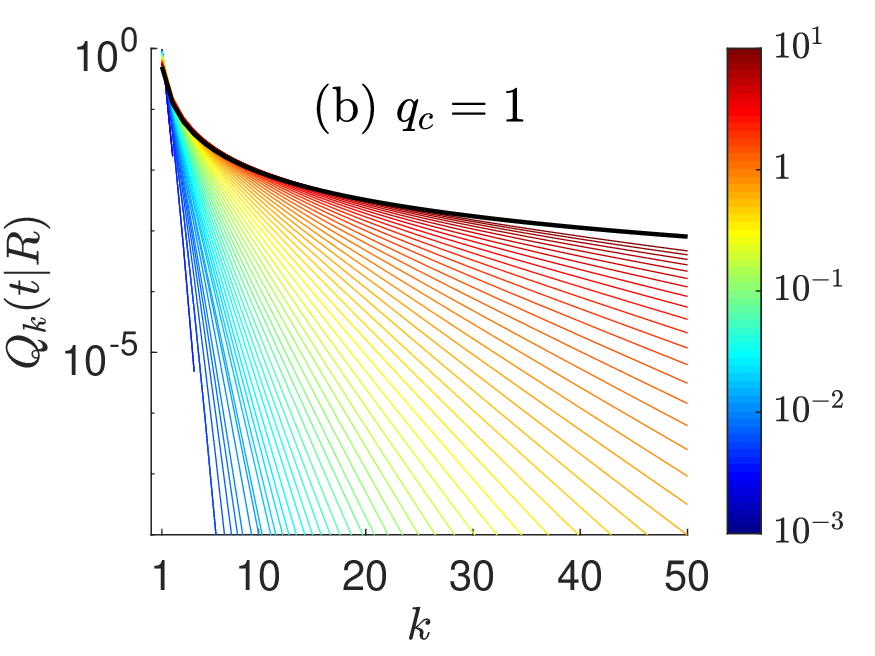} 
\includegraphics[width=0.32\textwidth]{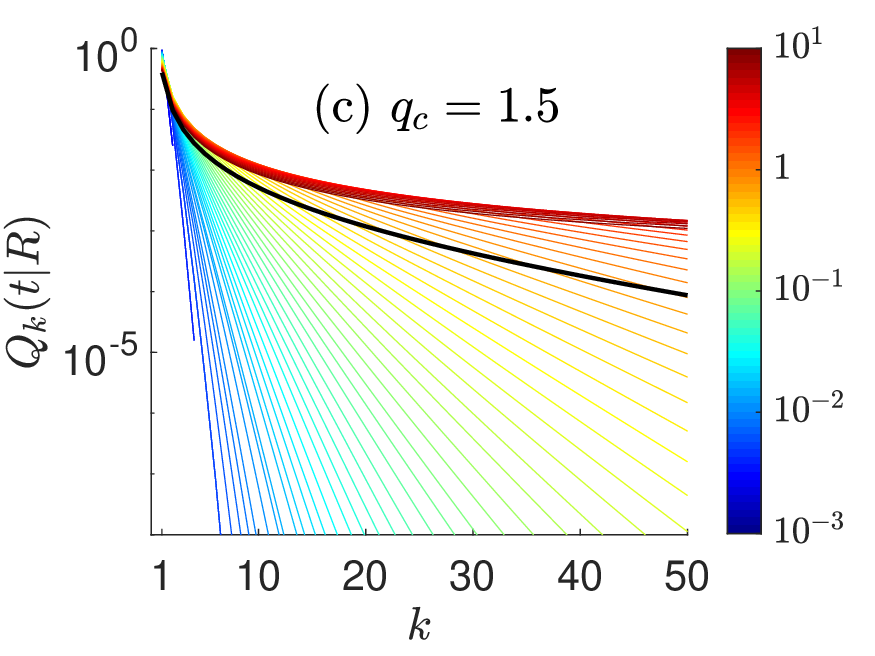} 
\end{center}
\caption{
Distribution $Q_k(t|R)$ of the population size for diffusion with $D =
1$ outside the unit ball ($R = 1$) in the subcritical ($\qc = 0.5$,
panel a), critical ($\qc = 1$, panel b), and supercritical ($\qc =
1.5$, panel c) regimes.  Solid black line indicates the steady-state
distribution $Q_k(\infty|R)$ given by Eq. (\ref{eq:Qk_asympt}),
whereas colored lines show 64 distributions with $t$ ranging on the
logarithmic scale from $10^{-3}$ (dark blue) to $10^1$ (dark red). }
\label{fig:Qk}
\end{figure}

\begin{figure}
\begin{center}
\includegraphics[width=0.32\textwidth]{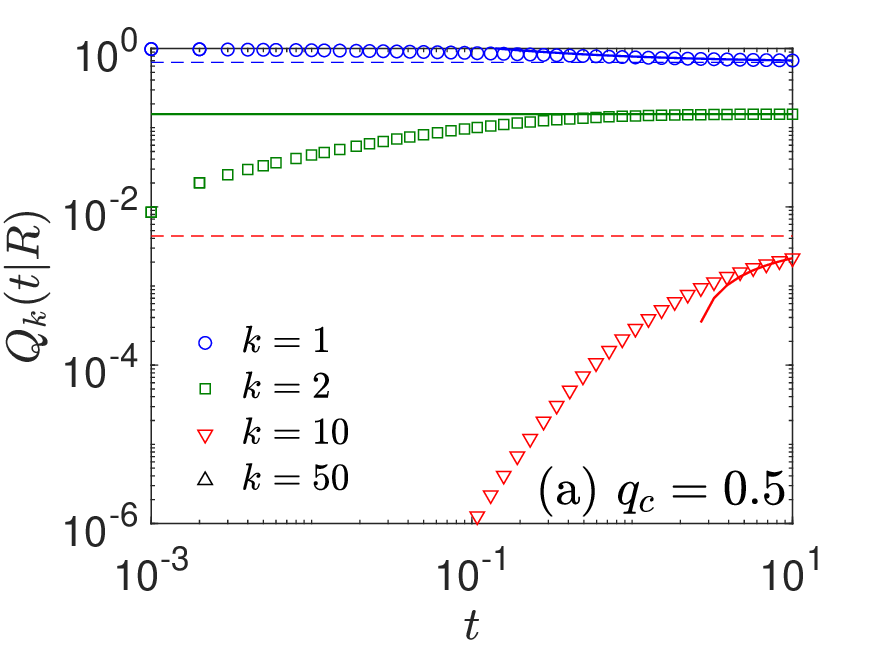} 
\includegraphics[width=0.32\textwidth]{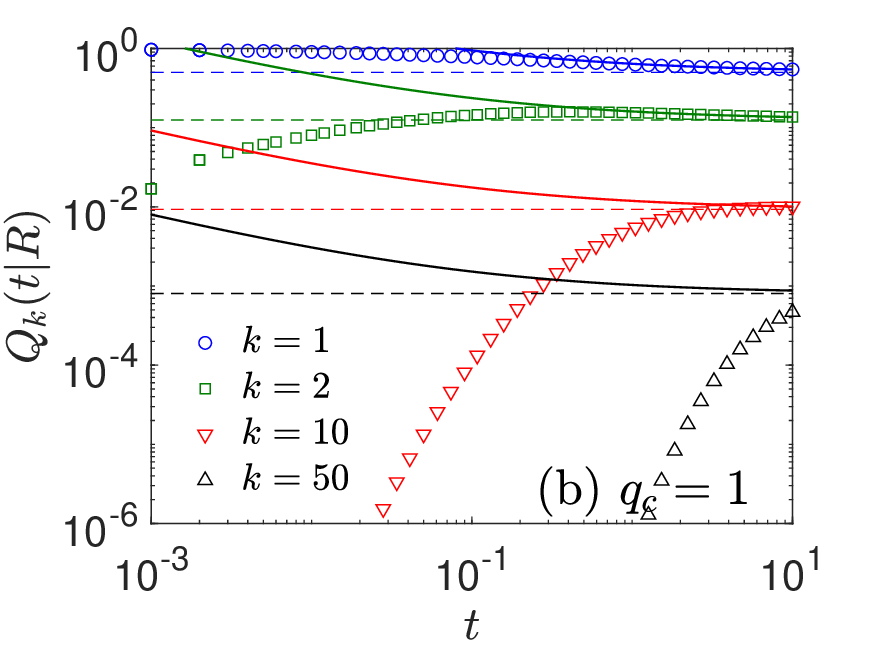} 
\includegraphics[width=0.32\textwidth]{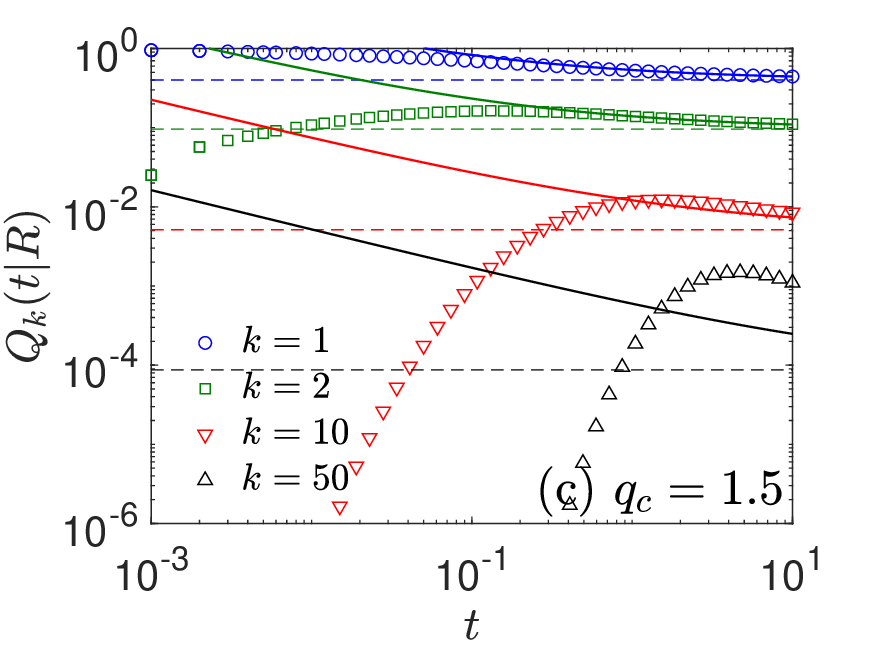} 
\end{center}
\caption{
Time evolution of the probabilities $Q_k(t|R)$ for diffusion with $D =
1$ outside the unit ball ($R = 1$) in the subcritical ($\qc = 0.5$,
panel a), critical ($\qc = 1$, panel b), and supercritical ($\qc =
1.5$, panel c) regimes.  Symbols present $Q_k(t|R)$ obtained via a
numerical integration of Eq. (\ref{eq:Qn}), solid lines show the
asymptotic relation (\ref{eq:Qk_long}), whereas dashed horizontal
lines indicate the corresponding steady-state values $Q_k(\infty|R)$
given by Eq. (\ref{eq:Qk_asympt}).  Note that the data for $k = 50$
are not visible on panel a, because $Q_{50}(t|R) < 10^{-6}$ in the
considered time interval.  Since $a_2 = 0$ at $\qc R = 0.5$, solid and
dashed curves coincide for $k = 2$ on panel a. }
\label{fig:Qk2}
\end{figure}

Figure \ref{fig:Qk} illustrates the time evolution of the distribution
$Q_k(t|R)$ in three regimes.  The shown curves are obtained via a
numerical integration of Eq. (\ref{eq:Qn}), see \ref{sec:numerics} for
details.
At the beginning ($t = 0$), a single particle is present in the system
so that $Q_k(0|R) = \delta_{k,1}$.  At short times, multiple branching
events are unlikely in all three regimes, yielding a sharp decrease of
$Q_k(t|R)$ with $k$.  As time increases, the distribution $Q_k(t|R)$
progressively approaches its steady-state limit $Q_k(\infty|R)$ given
by Eq. (\ref{eq:Qk_asympt}) and shown by solid black line.  However,
according to Eq. (\ref{eq:Qk_long}), this approach is rather slow and
in general not monotonous in time, as illustrated in
Fig. \ref{fig:Qk2}.  In fact, while the probability of no branching,
$Q_1(t|R)$, which represents the survival probability for conventional
diffusion-controlled reactions, decreases monotonically, the other
probabilities $Q_k(t|R)$ generally exhibit non-monotonous behavior.
This is particularly clear in the supercritical regime: at short
times, the probability of having $k\geq 2$ particles increases due to
branching events; however, at longer times, more and more particles
are present, and the likelihood of keeping a fixed number of particles
decreases.  A similar behavior can be observed in the critical and
subcritical regimes.  The leading-order correction in
Eq. (\ref{eq:Qk_long}) accurately captures the long-time behavior for
$k = 1$ and $k = 2$, whereas longer times are needed to observe their
agreement for $k = 50$.  In the critical regime, $a_k$ are positive,
and all leading-order asymptotic curves (\ref{eq:Qk_long}) are thus
decreasing.  This highlights the existence of the maxima of $Q_k(t|R)$
for $k \geq 2$, even though such a maximum is only visible for $k = 2$
at the considered range of times.

\subsection{Limiting population size distribution in the supercritical regime}
\label{sec:limiting_distribution}

In the supercritical regime ($\qc R > 1$), the mean population size
grows exponentially at long times, with the growth rate $\mu$ given by
Eq. (\ref{eq:mu_def}).  This motivates introducing the rescaled
population $\eta(t) = e^{-\mu t} \N(t)$.  Assuming that $\eta(t)$
converges in distribution to a nondegenerate random variable $\eta$,
we denote its Laplace transform by
\begin{equation}     \label{eq:Phi_def}
\Phi(z) = \E \{ e^{-z \eta}\} =\lim_{t\to\infty} G_{\exp(-z e^{-\mu t})}(t|R) ,
\end{equation}
with the starting point at the surface.  The integral equation
(\ref{eq:Gs_integral}) for the generating function $G_s(t|R)$ yields
an exact nonlinear equation for $\Phi(z)$.  Here, we proceed with a
more direct argument.  The initial particle started from the spherical
surface experiences either no branching with probability
$Q_1(\infty|R) = 1/(1+q_c R)$, or branches into two particles at a
random time $\tau$ with the probability density
\begin{equation}  \label{eq:Kdef}
K(t) = 4\pi R^2 \qc D \, P^+(R,t|R),  \qquad \int\limits_0^\infty dt \, K(t) = \gamma_+ < 1,
\end{equation}
with $P^+(R,t|R)$ given by Eq. (\ref{eq:P+R}), and $\gamma_+$ defined
in Eq. (\ref{eq:gamma_plus}).  In the second case, $\N(t) =
\N_1(t-\tau) + \N_2(t-\tau)$, where $\N_1$ and $\N_2$ are random sizes
of two independent subpopulations generated by the newborn particles
in the remaining time $t-\tau$.  As a consequence,
\begin{equation*}
\fl
\eta(t) = e^{-\mu \tau} \bigl[e^{-\mu(t-\tau)} \N_1(t-\tau) + e^{-\mu(t-\tau)} \N_2(t-\tau)\bigr] 
= e^{-\mu \tau} \bigl[\eta_1(t-\tau) + \eta_2(t-\tau)\bigr].
\end{equation*}
In the limit $t\to\infty$, one gets $\eta = e^{-\mu \tau} (\eta_1 +
\eta_2)$, where $\eta_1$ and $\eta_2$ are two independent copies of
$\eta$.  Conditioning on the first branching time gives thus
\begin{equation}   \label{eq:Phi_integral}
\Phi(z) = \frac{1}{1+q_c R} + \int\limits_0^\infty dt\,K(t) \left[\Phi(ze^{-\mu t})\right]^2 .
\end{equation}
This nonlinear equation characterizes the limiting population
distribution without requiring the time-dependent probabilities
$Q_k(t|R)$.

One easily checks that $\Phi(0)=1$.  Differentiating
Eq. (\ref{eq:Phi_integral}) at $z=0$, one recovers the recurrence
relation (\ref{eq:nk_recursive}) for the rescaled moments $m_k =
\mathbb{E}\{\eta^k\}$, see Eq. (\ref{eq:Nk_supercrit}).
The limiting rescaled population size $\eta$ may have an atom at zero,
whose weight $p_0 = \P\{\eta=0\} = \Phi(\infty)$ satisfies $p_0 =
1-\gamma_+ + \gamma_+ p_0^2$ according to Eq. (\ref{eq:Phi_integral}).
Besides the trivial solution $p_0=1$, this equation admits the
solution $p_0 = 1/(q_cR)$, which coincides with the probability of a
finite population obtained in Sec. \ref{sec:distrib}.

\begin{figure}
\begin{center}
\includegraphics[width=0.49\textwidth]{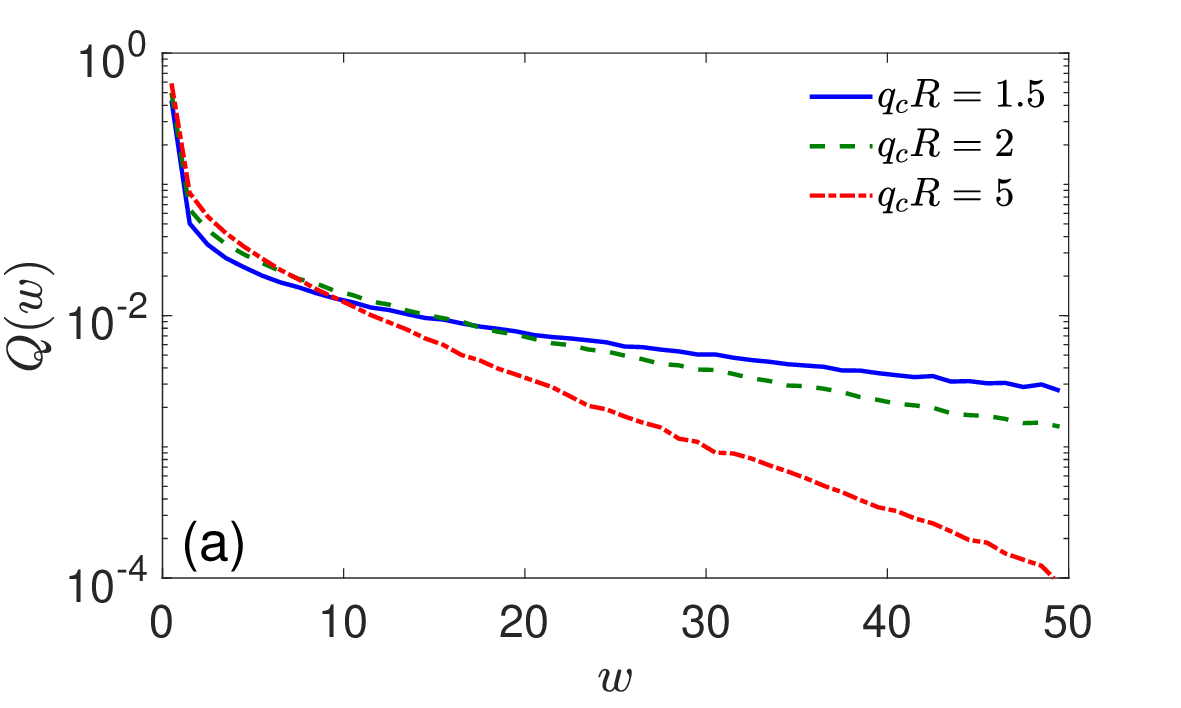} 
\includegraphics[width=0.49\textwidth]{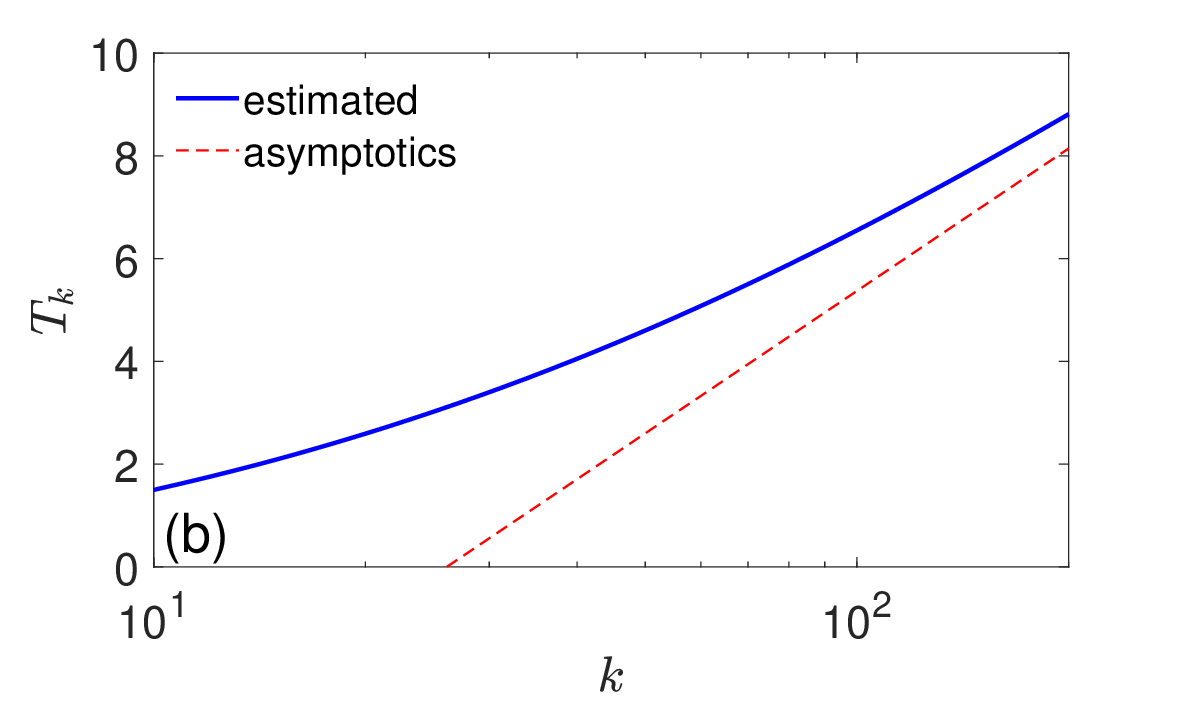} 
\end{center}
\caption{
{\bf (a)} The probability density $Q(w)$ of the rescaled limiting
population size $\bar{\eta} = (\eta \mid \eta > 0)$ in the
supercritical regime, for three values of $\qc R$ as indicated in the
legend (see \ref{sec:smoothing}).
{\bf (b)} The maximum $t_k$ of $Q_k(t|R)$ for $\qc R = 1.5$.  The
estimated value (solid line) is compared to the asymptotic relation
(\ref{eq:tmax}), with $w_* \approx 26.12$.}
\label{fig:Qlim}
%
\end{figure}

The probability density $Q(w)$ of the conditional variable $\bar{\eta}
= (\eta \mid \eta > 0)$ can be formally obtained via the inverse
Laplace transform of
\begin{equation}
\Phi(z) = \frac{1}{\qc R} + \biggl(1 - \frac{1}{\qc R}\biggr)\int\limits_0^\infty dw \, e^{-zw}\, Q(w),
\qquad \int\limits_0^\infty dw \, Q(w) = 1.
\end{equation}
However, its direct numerical computation is subtle: in order to
perform the inverse Laplace transform, one would need to solve
Eq. (\ref{eq:Phi_integral}) for complex-valued $z$.  As a systematic
analysis of this problem is beyond the scope of this paper, we used a
simpler Monte Carlo method described in \ref{sec:smoothing}.  Figure
\ref{fig:Qlim}a illustrates the probability density $Q(w)$ computed for 
several values of $\qc R$.

The probability density $Q(w)$ characterizes the growing part of the
population-size distribution.  In the joint limit $t\to\infty$ and
$k\to\infty$, with $w = k e^{-\mu t}$ fixed, one expects the scaling
form
\begin{equation} \label{eq:Qk_asymptQ}
Q_k(t|R) \simeq \biggl(1 - \frac{1}{\qc R}\biggr) e^{-\mu t} Q(e^{-\mu t} k) \qquad (t\to\infty).
\end{equation}
This approximation describes population sizes of order $e^{\mu t}$ and
should not be confused with the fixed-$k$ long-time limit obtained in
Sec. \ref{sec:distrib}.  

The scaling form (\ref{eq:Qk_asymptQ}) suggests a traveling-wave-like
propagation of the population-size distribution towards larger values
of $k$.  Indeed, for sufficiently large $k$, the time dependence of
$Q_k(t|R)$ within the scaling regime is governed by the function
$wQ(w)$, where $w = ke^{-\mu t}$.  If this function has a unique
maximum at $w = w_*$, determined by $Q(w_*) + w_* Q'(w_*)=0$, the
corresponding peak occurs approximately at
\begin{equation}  \label{eq:tmax}
t_k\approx \frac{\ln(k/w_*)}{\mu} \,.  
\end{equation}
Thus, the characteristic population size grows exponentially with
time, while the position of the peak propagates linearly on a
logarithmic population-size scale.  Figure \ref{fig:Qlim}b shows $t_k$
as a function of $k$ for $\qc R = 1.5$ and illustrates the trend
toward the predicted asymptotic behavior (\ref{eq:tmax}).  The value
of $w_* \approx 26.12$ was estimated numerically (see
\ref{sec:smoothing}).

\section{Mean population size in higher dimensions}
\label{sec:higher}

While we focused on the three-dimensional case, similar analysis can
be performed in higher dimensions.  As discussed in
\cite{Grebenkov26a}, the subcritical and supercritical regimes are
separated by the critical catalytic rate $\qccrit = (d-2)/R$.  In this
section, we restrict our discussion to the mean population size
$N_1(t|\x_0)$, which is obtained by integrating the propagator
$P^{-}(\x,t|\x_0)$ over $\x\in \Omega$ and thus obeys the standard
linear diffusion equation \cite{Grebenkov26a,Grebenkov26b}:
\numparts
\begin{eqnarray}
&& \partial_t N_1 - D\Delta N_1 = 0 \qquad (\x_0\in\Omega), \\ 
&& \partial_n N_1 - \qc N_1 = 0 \qquad (\x_0\in \pa), \\
&& N_1 \to 1 \qquad (|\x_0|\to\infty) ,\\
&& N_1(0|\x_0) = 1 .
\end{eqnarray}
\endnumparts
In turn, its Laplace transform $\tilde{N}_1(p|r_0) =
\int\limits_0^\infty dt \, e^{-pt} \, N_1(t|\x_0)$ satisfies
\numparts
\begin{eqnarray}
&& \biggl(p-D \biggl(\partial_{r_0}^2 + \frac{d-1}{r_0} \partial_{r_0}\biggr)\biggr) \tilde{N}_1  = 1 \qquad (R < r_0 < \infty), \\
&& -\partial_{r_0} \tilde{N}_1 - \qc \tilde{N}_1 = 0 \qquad (r_0 = R), \\
&& \tilde{N}_1  \to 1/p \qquad (r_0\to\infty),
\end{eqnarray}
\endnumparts
where we wrote the radial part of the Laplace operator in spherical
coordinates with $r_0 = |\x_0|$ and $-\partial_{r_0} = \partial_n$ for
the normal derivative oriented outward from $\Omega$.  Expectedly, the
problem does not include angular coordinates.  The solution can be
written as
\begin{equation}  \label{eq:tildeN1_d}
\fl
\tilde{N}_1(p|r_0) = \frac{1 + u_d(p|r_0)}{p} \,, \qquad 
u_d(p|r_0) = (r_0/R)^{-\nu}  \frac{\qc \, K_{\nu}(\alpha r_0)}{\alpha K_{\nu+1}(\alpha R) - \qc K_{\nu}(\alpha R)} \,,
\end{equation}
where $\alpha = \sqrt{p/D}$, $\nu = d/2 - 1$, $K_\nu(z)$ is the
modified Bessel function of the second kind, and the normalization
constant was obtained from the boundary condition.

In order to perform the Laplace transform inversion, we apply the
following identity
\begin{equation}
\L^{-1}_{s\to u} \left\{ \frac{K_\nu(\sqrt{s}\, \xi)}{\sqrt{s} K_{\nu+1}(\sqrt{s}) + \mu K_\nu(\sqrt{s})} \right\} 
= \frac{2}{\pi} \int\limits_0^\infty dz\, z\, e^{-z^2 u}\, g_{\nu}(z; \xi),
\end{equation}
with
\begin{equation}
g_\nu(z; \xi) = \frac{Y_\nu(z\xi)(z J_{\nu+1}(z) + \mu J_\nu(z)) - J_\nu(z\xi)(z Y_{\nu+1}(z) + \mu Y_\nu(z))}
{(z J_{\nu+1}(z) + \mu J_\nu(z))^2 + (z Y_{\nu+1}(z) + \mu Y_\nu(z))^2} \,,
\end{equation}   
which is valid for $\nu > - 1$, $\mu > -2\nu$, and $\xi \geq 1$.
Setting $\nu = d/2-1$, $\mu = -\qc R$, $\xi = r_0/R$, and $u =
Dt/R^2$, we get
\begin{equation}
N_1(t|r_0) = 1 + \frac{2\qc R}{\pi} (R/r_0)^{\frac{d}{2}-1} \int\limits_0^\infty \frac{dz}{z} \, (1 - e^{-z^2 Dt/R^2})\, 
g_{\frac{d}{2}-1}(z; r_0/R).
\end{equation} 
This expression allows for an accurate computation of the mean
population size for $\qc < \qccrit$ in any spatial dimension.

The analysis of the long-time behavior is more convenient in the
Laplace domain.  As the long-time limit corresponds to $p\to 0$, we
substitute $K_\nu(z) \simeq \Gamma(\nu) 2^{\nu-1} z^{-\nu}$ to
Eq. (\ref{eq:tildeN1_d}) to obtain in the leading order
\begin{equation}
u_d(p|r_0) \simeq \frac{(R/r_0)^{d-2}}{(d-2)/(\qc R) - 1}  \qquad (p\to 0).
\end{equation}
One sees that this limit is finite for any $\qc < \qccrit = (d-2)/R$.
As a consequence, we get in this regime
\begin{equation}
N_1(t|r_0) \to N_1(\infty|r_0) = 1 + \frac{(R/r_0)^{d-2}}{\qccrit/\qc - 1} > 1 \qquad (t\to \infty).
\end{equation}
In this subcritical regime, escape events dominate over catalytic
branching, so that the mean population size remains finite at long
times.  Although the initial particle and some of its offspring may
replicate upon encountering the catalytic sphere, each particle has a
nonzero probability of escaping to infinity without further
replication.  The farther the starting point is from the catalytic
surface (i.e., the larger $r_0$), the lower the probability that the
initial particle will reach the sphere and undergo replication.
Finally, note that $N_1(\infty|r_0)$ diverges as $\qc$ approaches the
critical value $\qccrit$ from below.

The behavior is different in the critical ($\qc = \qccrit$) and
supercritical ($\qc > \qccrit$) regimes, for which we need to keep
higher-order terms in $p$.  We sketch several results in the critical
regime:

(i) For $d = 4$, we get
\begin{equation}
u_4(p|r_0) \simeq \frac{4D}{p r_0^2 (\ln(\frac{4D}{pR^2}) - 2\gamma)}   \qquad (p\to 0)
\end{equation}
in the leading order, where $\gamma \simeq 0.5772$ is the Euler
constant.  The Tauberian theorem then implies in the leading order
\begin{equation}
N_1(t|r_0) \simeq \frac{4D t}{r_0^2 \ln(t/t_0)}   \qquad (t\to\infty),
\end{equation}
with $t_0 = e^{2\gamma} R^2/(4D) $.  This asymptotic behavior is
confirmed numerically (not shown).

(ii) For $d = 5$, we have
\begin{equation}  \label{eq:u5}
u_5(p|r_0) = \frac{\qc R^4 (\alpha r_0 + 1) e^{-\alpha(r_0-R)}}{r_0^3 (\alpha^2 R^2 + (\alpha R+1) (-\qc R + 3))} \,.
\end{equation}
In the critical regime ($\qc = 3/R$), one can explicitly invert the
Laplace transform to get
\begin{eqnarray} \nonumber
N_1(t|r_0) &=& 1 + \frac{3R}{r_0^3} \left\{\frac{(r_0+R)\sqrt{Dt}\, e^{-(r_0-R)^2/(4Dt)}}{\sqrt{\pi}} \right. \\
& +& \left. \erfc\left(\frac{r_0-R}{\sqrt{4Dt}}\right) \left(Dt - \frac{r_0^2-R^2}{2}\right)\right\} .
\end{eqnarray}
At long times, one can use $u_5(p|r_0) \sim 3RD/(r_0^3 p)$ as $p\to
0$ so that $\tilde{N}_1\simeq 3RD/(r_0^3 p^2)$ and thus $N_1(t|r_0)
\simeq 3RDt/r_0^3$.

(iii) More generally, for any $\nu > 1$ (i.e., $d \geq 5$), we can use
the asymptotic behavior
\begin{equation}
K_\nu(z) \simeq 2^{\nu-1} \Gamma(\nu) z^{-\nu} - 2^{\nu-3} \Gamma(\nu-1) z^{-\nu+2}   \qquad (z\to 0),
\end{equation}  
to get in the critical regime:
\begin{equation}
u_d(p|r_0) \simeq (R/r_0)^{2\nu} \frac{4\nu(\nu-1)}{\alpha^2 R^2}  \qquad (p\to 0)
\end{equation}
and thus
\begin{equation}
N_1(t|r_0) \simeq (R/r_0)^{d-2} \frac{(d-2)(d-4)}{R^2} Dt  \qquad (t\to\infty)
\end{equation}
in the leading order, for $d \geq 5$. 

We summarize these asymptotic results for the critical regime as
\begin{equation}  \label{eq:N1_d_asympt}
N_1(t|r_0) \propto \left\{ \begin{array}{c l} \sqrt{t} & (d = 3), \\
t/\ln(t) & (d = 4), \\
t  & (d \geq 5). \\ \end{array} \right.
\end{equation} 
The above asymptotic behavior of the mean population size is similar
to that found for lattice random walks with a single catalytic site
\cite{Albeverio98}.  This agreement suggests that the critical
growth exponents are determined primarily by the large-scale
properties of diffusion and are insensitive to microscopic details of
the catalytic mechanism.  In particular, the spherical catalytic
surface behaves effectively as a localized branching site at
sufficiently large length and time scales, although the prefactors and
subleading corrections retain information about the geometry.

Finally, the supercritical regime is characterized by an exponential
growth of the mean population size at long times: $N_1(t|r_0) \propto
e^{\nnu t}$.  In three dimensions, we have found the growth rate $\nnu
= D(\qc - 1/R)^2$, as well as the prefactor (Sec. \ref{sec:Nk}).  This
rate could alternatively be found as the largest pole of the function
$u_3(p|r_0)$.  In higher dimensions, the growth rate is similarly
determined by the largest pole of $\tilde{N}_1(p|r_0)$.  In fact, we
search for a positive zero of the function
\begin{equation}  \label{eq:z0_eq}
z K_{\nu+1}(z) - \qc R K_\nu(z) = 0
\end{equation}
or, equivalently, $f_\nu(z) = \qc R$, where $f_\nu(z) =
zK_{\nu+1}(z)/K_\nu(z)$.  The function $f_\nu(z)$ increases
monotonically from $f_\nu(0) = 2\nu$ to $f_\nu(\infty) =\infty$ as $z$
ranges from $0$ to infinity (see \ref{sec:BesselK}).  As a
consequence, if $\qc R > 2\nu = (d-2)$ (i.e., $\qc > \qccrit =
(d-2)/R$), the equation $f_\nu(z) = \qc R$ has a unique solution that
we denote as $z_0$.  This solution determines the growth rate $\nnu =
D z_0^2/R^2$.  For instance, the explicit form (\ref{eq:u5}) for $d =
5$ yields
\begin{equation}
\nnu = \frac{D}{R^2} \left(\frac{\qc R - 3 + \sqrt{(\qc R - 3)(\qc R + 1)}}{2}\right)^2.
\end{equation}
In two dimensions, Eq. (\ref{eq:z0_eq}) with $\nu = 0$ has the unique
solution $z_0$ for any $\qc > 0$ so that $N_1(t|\x_0)$ always exhibits
an exponential growth at long times.

\section{Steady-state population size distribution in general exterior domains}
\label{sec:generic_domains}

The explicit steady-state distribution derived above for the exterior
of a ball relies strongly on rotational invariance.  In particular,
when the initial particle is released from the spherical boundary, all
probabilities $Q_k(\infty|\x)$ are constant on the boundary, and the
integral equation reduces to the scalar recursion (\ref{eq:QnR}).
Nevertheless, the steady-state problem admits a useful operator
formulation for a generic exterior domain $\Omega =
\R^d\setminus\overline{\mathcal{O}}$ (with $d\geq 3$), i.e., the
exterior of a bounded set $\mathcal{O}$ with a sufficiently smooth
connected boundary $\pa$.  We assume that the whole boundary is
catalytic, with a constant catalytic rate $\qc > 0$.  In the limit
$t\to\infty$, the integral equation (\ref{eq:Qn_integral}) for $k \geq
2$ is reduced to
\begin{equation} \label{eq:Qn_integral_inf}
Q_k(\infty|\x_0) = \qc \int\limits_{\pa} d\x \, G^{+}(\x,\x_0) \, \sum\limits_{j=1}^{k-1} Q_j(\infty|\x) Q_{k-j}(\infty|\x) 
\quad (k\geq 2),
\end{equation}
where
\begin{equation}
G^+(\x,\x_0) = \int\limits_0^\infty dt \, D P^+(\x,t|\x_0)
\end{equation}
is the Robin Green's function.  As the right-hand side depends on
$Q_1(\infty|\x), \ldots, Q_{k-1}(\infty|\x)$, this equation determines
$Q_k(\infty|\x_0)$ recursively.  Since $Q_1(t|\x_0)$ is the (ordinary)
survival probability up to time $t$, its limit $Q_1(\infty|\x_0)$ is
the probability of escape to infinity, which satisfies
\begin{equation}  \label{eq:Q1inf}
\Delta Q_1(\infty|\x_0) = 0 \quad (\x_0\in\Omega), \qquad 
\left\{ \begin{array}{rl} \partial_n Q_1 + \qc Q_1 = 0 & \quad (\x_0\in\pa), \\
 Q_1(\infty|\x_0) \to 1 & \quad (|\x_0|\to \infty). \end{array} \right.
\end{equation}

To reveal the critical behavior of the steady-state population size
distribution, we restrict the starting point $\x_0$ to the boundary
and introduce the exterior Dirichlet-to-Neumann (DtN) operator $\M$:
for a prescribed function $f$ on $\pa$, we define $\M f =
\left.\partial_n u\right|_{\pa}$, where $u$ is the unique solution of
\begin{equation}
\Delta u =0 \quad (\x\in\Omega),\qquad  u = f \quad(\x\in\pa), \qquad u \to0 \quad (|\x|\to\infty).
\end{equation}
With this convention, $\M$ is a positive self-adjoint operator with a
discrete spectrum \cite{Arendt15,Bundrock26}.  Its eigenvalues can be
arranged by $n = 0,1,2,\ldots$ in nondecreasing order: $0 < \mu_0 <
\mu_1 \leq \mu_2 \leq \ldots \nearrow +\infty$.  The principal
(lowest) eigenvalue $\mu_0$ is strictly positive and simple.  For the
exterior of a ball of radius $R$, the constant function is the
principal eigenfunction and $\mu_0 = (d-2)/R$.  A central role is
played by the boundary resolvent operator
\begin{equation}     \label{eq:Koperator}
\K = \qc (\M + q_c I)^{-1}.
\end{equation}
For a spherical surface in $\R^3$, $\K$ acting on a constant function
reduces to multiplication by $\gamma_+$ introduced in
Eq. (\ref{eq:gamma_plus}).

The following analysis remains formal and requires rigorous proofs
that are beyond the scope of the paper.  For brevity, let $q_k(\x) =
Q_k(\infty|\x)$ denote the probability of having a finite population
of $k$ particles at $t=\infty$, when the initial particle starts from
the boundary point $\x\in \pa$.
The restriction of the solution of Eq. (\ref{eq:Q1inf}) to $\pa$ can
be written as
\begin{equation}      \label{eq:q1_generic}
q_1 = 1 - \K 1 = (I - \K)1 = (\M + \qc I)^{-1} \M 1.
\end{equation}
In turn, using the spectral expansion of the Green's function
$G^+(\x,\x_0)$ on the boundary in terms of the DtN eigenfunctions, one
can interpret $\qc G^+(\x,\x_0)$ as the integral kernel of the
operator $\K$; consequently, Eq. (\ref{eq:Qn_integral_inf}) yields the
nonlinear recursion
\begin{equation}      \label{eq:qk_generic_recursion}
q_k = \K \left[\sum\limits_{j=1}^{k-1} q_j q_{k-j}\right]  \qquad (k\geq2).
\end{equation}
Equation (\ref{eq:qk_generic_recursion}) can be viewed as an
operator-valued generalization of the Catalan recursion
(\ref{eq:Qk_relation}).  For a sphere in $\R^3$, all $q_k$ are
constant on the boundary, and $\K$ reduces to multiplication by
$\gamma_+$.  In turn, for a generic domain, the pointwise products in
Eq. (\ref{eq:qk_generic_recursion}) couple different DtN
eigenfunctions, so that the distribution is generally controlled by
the whole DtN operator rather than by its lowest eigenvalue alone.

The operator representation provides a direct geometric
characterization of the transition between the subcritical and
supercritical regimes.  Each branching event produces two descendants,
whose subsequent branching locations are governed by the boundary
operator $\Gamma$.  The corresponding mean offspring operator is
therefore $2\Gamma$.  The critical point is reached when its spectral
radius equals one: $\rho(2\Gamma)=1$.  Since the spectral radius of
$\K$ is determined by the lowest eigenvalue $\mu_0$, $\rho(\K) =
\qc/(\qc + \mu_0)$, one recovers the critical catalytic rate
$\qccrit = \mu_0$ obtained in \cite{Grebenkov26a}.  For a sphere,
$\mu_0 = (d-2)/R$, and we retrieve the critical value $\qccrit =
(d-2)/R$ obtained above.

The operator formulation also provides insight into the large-$k$
behavior of the steady-state probabilities.  Since the coefficients
$q_k(\x)$ are nonnegative, their asymptotic behavior is governed by
the first positive singularity $s_*$ of the generating function
\begin{equation}      \label{eq:gs_generic}
g_s(\x) = \sum\limits_{k=1}^{\infty} q_k(\x)\, s^k \qquad (\x\in\pa).
\end{equation}
Summing Eq. (\ref{eq:qk_generic_recursion}) over $k$ yields the
nonlinear operator equation $g_s = s q_1 + \K g_s^2 $ or,
equivalently,
\begin{equation}    \label{eq:gs_DtN_compact}
\M(g_s-s) = \qc g_s(g_s-1).
\end{equation}
Thus, the steady-state population problem in a generic exterior domain
is reduced to a nonlinear equation on its boundary.

For a spherical surface, the nonlinear equation
(\ref{eq:gs_DtN_compact}) reduces to a quadratic equation whose
solution exhibits a square-root singularity at $s = s_*$.  This
singularity is responsible for the asymptotic behavior
(\ref{eq:Qkinf_largek}).  Although Eq. (\ref{eq:gs_DtN_compact}) is no
longer a scalar quadratic equation for a generic boundary, its
nonlinearity retains the same quadratic structure.  It is therefore
natural to expect a similar square-root singularity of the generating
function $g_s(\x)$, at a geometry-dependent value $s = s_*$.
At the critical point $q_c=\mu_0$, one can further argue that $s_*=1$.
Indeed, the derivative $h(\x)= \partial_s g_s(\x)|_{s=1}$ represents
the mean population size.  Assuming that this mean is finite,
differentiating Eq. (\ref{eq:gs_DtN_compact}) at $s=1$, where $g_1=1$,
yields $(\M-\mu_0 I) h = \M 1$.  However, the operator on the left has
a zero eigenvalue, whereas the right-hand side has a strictly positive
projection onto its principal eigenfunction.  This contradiction
implies that the mean population diverges at criticality, so that
$g_s$ is nonanalytic at $s=1$.

If this singularity is indeed of square-root type and dominates the
large-order coefficients of the generating function, the steady-state
probabilities are expected to obey
\begin{equation}
q_k(\x) \sim A(\x)\, k^{-3/2} \, s_*^{-k} \qquad (k\to\infty),
\end{equation}
where $A(\x)$ is a positive amplitude that depends on the starting
point and the boundary geometry.  This form encompasses both
subcritical and supercritical regimes, for which $s_*>1$ is expected
to produce an exponential cutoff.  In the supercritical regime, the
asymptotic expression describes only the finite-population part of the
distribution because unbounded population growth has a nonzero
probability.  At the critical point $q_c = \mu_0$, the singularity at
$s_* = 1$ eliminates the exponential cutoff, leaving the expected
power-law decay $q_k(\x)\propto k^{-3/2}$.

Thus, the exact spherical results suggest that the exponent $3/2$ is a
robust feature of surface-mediated branching, originating from the
quadratic structure of the generating-function equation.  In contrast,
the exponential decay rate $\ln s_*$ and the amplitude $A(\x)$ encode
the geometry of the catalytic boundary.  Establishing this asymptotic
picture rigorously for general exterior domains remains an open
problem.

\section{Conclusion}
\label{sec:conclusion}

In this paper, we investigated diffusion-driven autocatalytic
branching at the spherical surface.  The rotational invariance of the
exterior domain and the explicit single-particle propagators allowed
us to characterize the statistics of the population size.  In three
dimensions, we derived explicit expressions for the mean population
size $N_1(t|\x_0)$ and the probability $Q_1(t|\x_0)$ of no branching,
while higher-order moments $N_k(t|\x_0)$ and probabilities
$Q_k(t|\x_0)$ were determined recursively from convolution-type
equations.  The long-time behavior of the moments exhibits three
regimes, separated by the critical catalytic rate $\qccrit = 1/R$:
finite limiting moments in the subcritical regime, power-law growth
$N_k(t|\x_0)\propto t^{k-1/2}$ at criticality, and exponential growth
$N_k(t|\x_0)\propto e^{k\mu t}$ in the supercritical regime, with
$\mu=D(q_c-1/R)^2$.
In contrast, the probabilities $Q_k(t|\x_0)$ approach finite
steady-state limits in all three regimes.  We obtained these limits
explicitly and showed that their large-$k$ behavior is governed by a
universal algebraic prefactor $k^{-3/2}$, accompanied by an
exponential cutoff $e^{-\eta k}$, where $\eta =
\ln[(1+q_cR)^2/(4q_cR)]$.  The exponential cutoff disappears at
criticality, leaving a pure power-law decay.  In the supercritical
regime, the limiting distribution over finite population sizes is not
normalized to unity, as unbounded population growth occurs with
nonzero probability.  We also investigated the limiting distribution
of the exponentially rescaled population $e^{-\mu t}\N(t)$,
characterized by a nonlinear integral equation and reconstructed
numerically.

We extended the analysis of the mean population size to spherical
surfaces in higher spatial dimensions.  At criticality, the
square-root growth found in three dimensions is replaced by $t/\ln t$
in four dimensions and by linear growth in higher dimensions.  These
scaling laws agree with those known for lattice random walks with a
single catalytic site, suggesting that they are governed by the
large-scale properties of diffusion rather than by the microscopic
details of the catalytic mechanism.

Finally, we considered catalytic branching in general exterior domains
through the exterior Dirichlet-to-Neumann operator.  This formulation
identifies the critical catalytic rate with the smallest eigenvalue of
this operator, $\qccrit = \mu_0$, and reduces the steady-state
population problem to a nonlinear boundary equation.  The quadratic
structure of this equation suggests that the $k^{-3/2}$
large-population prefactor persists for generic catalytic surfaces,
while the exponential cutoff and its amplitude depend on the boundary
geometry.  Establishing these asymptotic predictions rigorously, as
well as extending the time-dependent analysis beyond spherical
symmetry, remains an interesting direction for future research.

\section*{Acknowledgements}
The author acknowledges the Simons Foundation for supporting his
sabbatical sojourn in 2024 at the CRM (CNRS -- Universit\'e de
Montr\'eal, Canada), as well as the Alexander von Humboldt Foundation
for support within a Bessel Prize award.  At the revision stage,
OpenAI's ChatGPT (GPT-5.6) was used to assist with language editing
and verifications.

\appendix
\section{Numerical computation}
\label{sec:numerics}

For a numerical computation of $N_k(t|\x_0)$ and $Q_k(t|\x_0)$ with $k
\geq 2$, one needs to solve the convolution-type equations
(\ref{eq:Nk_intR}) and (\ref{eq:Qn}), respectively.  As these
equations are linear with respect to $N_k(t|R)$ (or $Q_k(t|R)$), one
can solve them recurrently for $k = 2,3,\ldots$.  Despite this
linearity, the presence of inhomogeneous terms involving products of
lower-order solutions makes a direct Laplace-transform approach less
convenient than time-domain integration.  Another practical difficulty
is the weak singularity of the kernels: $P^{\pm}(R,t|R) \propto
t^{-1/2}$ as $t\to 0$.  This issue can be resolved by rescaling the
kernel and approximating a convolution as
\begin{equation} \label{eq:conv_integral}
\int\limits_0^t \frac{dt'}{\sqrt{t'}} \, p(t') \, f(t-t') \approx \sum\limits_{j=0}^{n} w_j \, f(t-j\delta),
\end{equation}
where $\delta$ is a chosen discretization timestep, $n = [t/\delta]$,
$f(t)$ denotes either $F_k(t|R)$ or $H_k(t|R)$,
\begin{equation}
p(t) = 4\pi R^2 \qc D \, P^{\pm}(R,t|R)\, \sqrt{t} 
\end{equation}
is the regularized kernel such that $p(0) = \qc \sqrt{D/\pi}$, and
$w_j$ are suitable quadrature weights.  For illustrative purposes, we
used the most basic, lowest-order quadrature to evaluate the integral
in Eq. (\ref{eq:conv_integral}), for which
\begin{equation}  \label{eq:wj}
\fl 
w_j = \frac{a_j + a_{j-1}}{2},  \qquad
a_j = \left(\sqrt{t_{j+1}} - \sqrt{t_j}\right) \bigl[p(j\delta) + p((j+1)\delta)\bigr]  
  \quad (j=0,1,\ldots,n),
\end{equation}
with $a_{-1} = a_n = 0$ (see more details in \cite{Grebenkov26c}).  In
all numerical examples, we set $R = D =1$ and used the timestep
$\delta = 10^{-3}$.

\section{Long-time asymptotic behavior}
\label{sec:Qk_long}

In this Appendix, we briefly discuss the asymptotic approach of
$Q_k(t|R)$ to its limit $Q_k(\infty|R)$ at long times in three
dimensions.  Setting $r_0 = R$ for simplicity, we first note that
\begin{equation}  \label{eq:Q1R}
Q_1(t|R) = Q_1(\infty|R) + (1- Q_1(\infty|R)) \erfcx\left(\sqrt{Dt}(\qc + 1/R)\right),
\end{equation}
so that
\begin{equation}
Q_1(t|R) - Q_1(\infty|R) \simeq \frac{1- Q_1(\infty|R)}{\sqrt{\pi D} (\qc + 1/R)}\, t^{-1/2} 
\qquad (t\to\infty),
\end{equation}
in the leading order.  A similar power-law approach is generally valid
for other probabilities $Q_k(t|R)$, as we show below.

Writing $Q_k(t|R) = Q_k(\infty|R) + A_k(t)$ and substituting it into
Eq. (\ref{eq:Qn}), we get
\begin{eqnarray*}
\fl
&& Q_k(\infty|R) + A_k(t) = 4\pi R^2 \qc D \int\limits_0^t dt' \, P^+(R,t'|R) \\
\fl
&& \times \sum\limits_{j=1}^{k-1}
\biggl( Q_j(\infty|R) Q_{k-j}(\infty|R) + 2 Q_{k-j}(\infty|R) A_j(t-t') + A_j(t-t') A_{k-j}(t-t')\biggr) \\
\fl
&& \approx Q_k(\infty|R) \biggl[1 - \erfcx\bigl(h_+ \sqrt{Dt}/R\bigr)\biggr]\\
\fl
&& + 8\pi R^2 \qc D\sum\limits_{j=1}^{k-1} Q_{k-j}(\infty|R) \int\limits_0^t dt' \, P^+(R,t'|R) A_j(t-t'),
\end{eqnarray*}
where $h_+ = 1 + \qc R$, and we used Eq. (\ref{eq:Pint}) to evaluate
the first term and neglected the last term, which is quadratic in $A$
and is thus smallen that the linear one.  The above linearized
equation can be solved in the Laplace domain as
\begin{equation}
\tilde{A}_k(p) \approx - \frac{Q_k(\infty|R)}{p + h_+ \sqrt{pD}/R} +
2 \sum\limits_{j=1}^{k-1} Q_{k-j}(\infty|R) \frac{\qc R \, \tilde{A_j}(p)}{1 + \qc R + R \sqrt{p/D}} \,,
\end{equation}  
where we used Eq. (\ref{eq:Pp_Laplace}) and the Laplace transform:
\begin{equation}
\int\limits_0^\infty dt \, e^{-pt} \, \erfcx(b\sqrt{t}) = \frac{1}{\sqrt{p}(b + \sqrt{p})}  \quad (b>0).
\end{equation}
As the long-time behavior corresponds to $p\to 0$, we have in the
leading order in $p$
\begin{equation}  \label{eq:tildeA_k}
\tilde{A}_k(p) \approx - \frac{Q_k(\infty|R) R}{h_+ \sqrt{pD}} +
2\gamma_+ \sum\limits_{j=1}^{k-1} Q_{k-j}(\infty|R) \tilde{A_j}(p)  \quad (p\to 0),
\end{equation}
where $\gamma_+ = \qc R/h_+ = 1-Q_1(\infty|R)$.  In addition, we rewrite
Eq. (\ref{eq:Q1R}) as
\begin{equation*}
A_1(t) = (1-Q_1(\infty|R))\, \erfcx\bigl(\sqrt{Dt}(\qc + 1/R)\bigr),
\end{equation*}
so that
\begin{equation}
\tilde{A}_1(p) = \frac{1-Q_1(\infty|R)}{p + \sqrt{pD} (\qc + 1/R)}  \approx \frac{\gamma_+}{h_+} \frac{R}{\sqrt{pD}}  \quad (p\to 0).
\end{equation}
Using this relation and the induction argument in
Eq. (\ref{eq:tildeA_k}), we obtain the leading small-$p$ behavior
\begin{equation}
\tilde{A}_k(p) \approx a_k \frac{R}{\sqrt{pD}}   \quad (p\to 0),
\end{equation}
and thus
\begin{equation}
A_k(t) \approx a_k \frac{R}{\sqrt{\pi Dt}}   \quad (t\to \infty),
\end{equation}
where the dimensionless coefficient $a_k$ is obtained via the
recurrent relation:
\begin{equation} 
\fl
a_1 = \frac{\gamma_+}{h_+}, \qquad 
a_k = - \frac{1-\gamma_+}{h_+} + 2\gamma_+ \sum\limits_{j=1}^{k-1} Q_{k-j}(\infty|R) \, a_j \quad (k=2,3,\ldots).
\end{equation}
Multiplying this relation by $z^k$, summing over $k\geq 2$, adding the
term $a_1 z$ to both sides, and using Eq. (\ref{eq:qz_def}), we get
the generating function:
\begin{equation}
a(z) = \sum\limits_{k=1}^\infty a_k z^k = \frac{1-\gamma_+}{2\gamma_+} 
\left(1 + \frac{2\gamma_+ z - 1}{\sqrt{1 - 4z\gamma_+(1-\gamma_+)}}\right).
\end{equation}
A lengthy computation of the $k$-th order derivative at $z = 0$ yields
\begin{equation}  \label{eq:ak_explicit0}
a_k = C_{k-1} \frac{(\qc R)^{k-1}}{(1+\qc R)^{2k}} \biggl[k \qc R - (k-1)\biggr] \qquad (k=1,2,\ldots).
\end{equation}
Substitution of the Catalan numbers $C_{k-1}$ from
Eq. (\ref{eq:Catalan}) gives Eq. (\ref{eq:ak_explicit}).

\section{Some properties of modified Bessel functions}
\label{sec:BesselK}

In this Appendix, we show that the function
\begin{equation}
f_\nu(z) = z \frac{K_{\nu+1}(z)}{K_\nu(z)}  \qquad (z > 0,~ \nu \geq 0)
\end{equation}
increases monotonically from $f_\nu(0) = 2\nu$ to $f_\nu(\infty) =
\infty$ as $z$ goes from $0$ to infinity (we expect that this property
is known but we could not find a reference).

The limit $f_\nu(0) = 2\nu$ follows immediately from the asymptotic
behavior of modified Bessel functions of the second kind:
$K_\nu(z)\simeq \Gamma(\nu) 2^{\nu-1} z^{-\nu}$ for $\nu > 0$ (in the
special case $\nu = 0$, $K_0(z)$ diverges logarithmically, yielding
$f_0(0) = 0$).
In turn, the limit $f_\nu(\infty) = \infty$ follows from the
asymptotic decay of $K_\nu(z)$.

In order to justify the monotonous increase, we evaluate the
derivative of $f_\nu(z)$:
\begin{equation}
z f'_\nu(z) = f_\nu^2(z) - 2\nu f_\nu(z) - z^2 = (f_\nu(z) - \nu)^2 - (z^2 + \nu^2),
\end{equation}
where we used the recurrence relations $K'_\nu(z) = \nu K_\nu(z)/z -
K_{\nu+1}(z)$ and $K'_{\nu+1}(z) = -K_\nu(z) - (\nu+1)K_{\nu+1}(z)/z$.
To proceed, we apply the inequality (see, e.g., \cite{Baricz09})
\begin{equation}  \label{eq:K_ineq}
z \frac{K'_\nu(z)}{K_\nu(z)} < - \sqrt{z^2 + \nu^2}  \qquad (z > 0, ~ \nu \in \R)
\end{equation}
that yields, together with $K_\nu(z) \geq 0$,
\begin{equation} \label{eq:auxil1}  
f_\nu(z) > \nu + \sqrt{z^2+\nu^2} ,
\end{equation}
from which $f'_\nu(z) > 0$, so that the function $f_\nu(z)$ grows
monotonically with $z$.

\section{Numerical reconstruction of the limiting distribution}
\label{sec:smoothing}

Under the assumption of convergence, the limiting distribution of the
rescaled population $\eta = \lim_{t\to\infty}e^{-\mu t}\N(t)$ can be
determined numerically by a population dynamics algorithm based on the
smoothing transform.  In the supercritical regime, $a=\qc R>1$, the
distribution of $\eta$ has an atom at zero of weight $1/a$.  To avoid
sampling this atom, we consider the conditional random variable
$\bar{\eta} = (\eta \mid \eta>0)$, whose probability density $Q(w)$
determines the continuous part of the distribution of $\eta$ as
\begin{equation}
Q_\eta(w) = (1 - 1/a) Q(w),  \qquad w>0.
\end{equation}
Conditioning on the first branching event and on the survival of at
least one descendant lineage yields the distributional identity
\begin{equation}    \label{eq:smoothing}
\bar{\eta} =_d e^{-U} \bigl(\bar{\eta}_1 + B \bar{\eta}_2 \bigr),
\end{equation}
where $\bar{\eta}_1$ and $\bar{\eta}_2$ are independent copies of
$\bar{\eta}$, and $B$ is an independent Bernoulli random variable with
$\mathbb{P}\{ B=1 \} = b = (a-1)/(a+1)$.  According to
Eqs. (\ref{eq:Pp_Laplace}) and (\ref{eq:Kdef}), the first-branching
time $\tau$ has the Laplace transform 
\begin{equation}
\E\{ e^{-p\tau}\} = \int\limits_0^\infty dt\, e^{-pt} \, K(t) = \frac{\qc R}{1 + \qc R} \, \frac{1}{1 + R \sqrt{p/D}/(1+\qc R)} \,.
\end{equation}
Consequently, the dimensionless first-branching time $U = \mu \tau$,
conditioned on branching, has the Laplace transform
\begin{equation}  \label{eq:Laplace_U}
\E\{ e^{-sU}\} = \frac{1}{1+b\sqrt{s}} \,, \qquad s = p/\mu.
\end{equation}
It can be sampled exactly as
\begin{equation}  \label{eq:U_generate}
U =_d \frac{b^2 E^2}{2Z^2} \,,
\end{equation}
where $E$ and $Z$ are independent standard exponential and Gaussian
random variables, respectively.  This representation follows from the
identity
\begin{equation}
\E_Z\left\{ \exp\left(-\frac{c^2}{2Z^2}\right) \right\} = 
\int\limits_{-\infty}^\infty dz \, \frac{e^{-z^2/2}}{\sqrt{2\pi}} \, e^{-c^2/(2z^2)} = e^{-c} \qquad (c\geq 0),
\end{equation}
where $Z$ is a standard Gaussian random variable.  Then, conditioning
on $E$ and averaging over its standard exponential distribution, one
gets
\begin{equation}
\E \{ e^{-sU} \} = \E_E\{ e^{-bE\sqrt{s}} \} = \int\limits_0^\infty dx\, e^{-x(1+b\sqrt{s})} = \frac{1}{1+b\sqrt{s}} \,.
\end{equation}

To construct the distribution of $\bar{\eta}$, we start from a large
ensemble of identical samples with the prescribed mean
\begin{equation}  \label{eq:bareta_mean}
\E\{\bar{\eta} \} = \frac{2a^2}{(a-1)^2} \,.
\end{equation}
At each iteration, every sample is replaced by $e^{-U}(\bar{\eta}_1 +
B \bar{\eta}_2)$, where $\bar{\eta}_1$ and $\bar{\eta}_2$ are drawn
independently from the preceding ensemble, while $U$ and $B$ are
generated independently.  After each iteration, we rescale all samples
by a common factor to restore the prescribed ensemble mean
(\ref{eq:bareta_mean}), thereby suppressing finite-sample fluctuations
of the overall scale.  The resulting ensemble approaches the
fixed-point distribution of Eq.~(\ref{eq:smoothing}).  The iteration
index is an auxiliary numerical parameter and does not represent
physical time.  The probability density is then estimated from the
converged ensemble, with the known weight $1-1/a$ restoring the
unconditional density.

Finally, the solution $w_*$ of the equation $Q(w_*) + w_* Q'(w_*) = 0$
can be obtained directly from the empirical data.  In fact, the
probability density of $Y = \ln \bar{\eta}$ is $p_Y(y) = e^y Q(e^y) =
w Q(w)$ so that $w_*$ corresponds to the maximum of $p_Y(y)$.
Computing an empirical density of $\ln(\bar{\eta})$ and finding the
location $y_*$ of its maximum yield $w_* = e^{y_*}$.

\end{document}